\documentclass[12pt,preprint]{aastex}

\slugcomment{Not to appear in Nonlearned J., 45.}

\shorttitle{High temperature combustion: Approaching equilibrium using
nuclear networks}
\shortauthors{Cabez\'on, Garc\'\i a-Senz and Bravo}

\begin{document}


\title{High temperature combustion: Approaching equilibrium \\
    using nuclear networks}


\author{R.M.Cabez\'on, D. Garc\'\i a-Senz\altaffilmark{1}
   and E. Bravo\altaffilmark{1}}
\affil{Departament de F\'\i sica i Enginyeria Nuclear, UPC, Jordi Girona 3, 
 M\`odul B5, 
\qquad\qquad    08034 Barcelona, Spain}

\email{ruben.cabezon@upc.es\qquad domingo.garcia@upc.es\qquad 
eduardo.bravo@upc.es}


\altaffiltext{1}{Institut D'Estudis Espacials de Catalunya, Gran Capit\`a 2-4 
08034 Barcelona}


\begin{abstract}
A method for integrating the chemical equations associated with nuclear
combustion at high temperature is presented and extensively checked. Following
the idea of M\"uller \citep{m86}, the feedback between nuclear rates and
temperature was taken into account by simultaneously computing molar
fraction changes and temperature response in the same matrix. The resulting
algorithm is very stable and efficient at calculating nuclear combustion in
explosive scenarios, especially in those situations where the reacting material
manages to climb to the nuclear statistical equilibrium regime. The numerical 
scheme may be useful not only for those who carry out hydrodynamical
simulations of explosive events, but also as a tool to investigate the
properties of a nuclear system approaching equilibrium through a variety of 
thermodynamical trajectories.

\end{abstract}




\section{Introduction}

Explosive phenomena in stars are often associated with the rapid release of
nuclear energy and a lavish production of nuclear species. In any calculation
dealing with novae or
supernovae explosions it is
highly desirable to  implement the nuclear part of hydrodynamic codes as
 better as
possible in order to get a suitable representation of these explosions and
nucleosynthetic yields. A
complication arises because of the huge quantity of nuclei involved,
which makes it difficult to simultaneously handle hydrodynamics and
nuclear combustion.
Recently, however, there have been performed 
hydrodynamic simulations in spherical symmetry incorporating 
several thousand nuclei (Rauscher et al. 2001). Also,
 calculations in two or three dimensions of both kinds of supernovae are quite
common. In these multidimensional calculations the mission of the nuclear
network is to provide a reasonable  nuclear energy rate at a minimum
computational cost. This is usually accomplished by using small networks of 7
or 13 isotopes (Timmes, Hoffman \& Woosley~2000). Detailed nucleosynthesis is
calculatedted {\it a
posteriori} by using a postprocessing technique with a larger network.

The most common procedure used for integrating the nuclear
networks handles the discretized chemical equations in their implicit form.
This enhances the stability of the nuclear system, which is very rigid
(Arnett 1996), allowing it to take longer time-steps. It is worthy of
note
that despite the great dependence of the rates on the temperature,
the effect of the latter is rarely incorporated implicitly into  numerical
codes. This
enforces to make a reliable extrapolation of the temperature and to take
small time-steps to guarantee stability, especially
when the nuclei approach the nuclear statistical equilibrium (NSE) regime.
Several methods have been suggested for handling  nuclear combustion
around the NSE regime. The most simple and fastest way is to use a conventional
integration method when the nuclear system is not too close to equilibrium
and to rely on statistical techniques to treat the phase in which direct
and reverse reactions are almost or completely balanced. Nevertheless, this
leads to
the problem of how to make a smooth transition between the two regimes.
Another possibility is to always use a standard integration method, which
takes
care of adequately restricting the time-step near equilibrium. In this case the
computational effort increases. There are also mixed schemes
(Hix \& Thielemann~1996)
in which reduced networks are used to describe the evolution of
light and a few intermediate-mass elements
(generally below $^{28}\mathrm {Si}$~or
$^{32}\mathrm{S}$) and a  
 statistical approach is taken for handling the remaining nuclei. While this
method is a very promising tool for explosive nucleosynthesis studies,
due to  its ability to handle hundreds of nuclei with a low computational
burden,
the decoupling between the thermal evolution and nuclear kinetics again
imposes a constraint on time-steps near NSE.

The importance of solving the nuclear network by coupling molar fractions and
temperature in the same matrix
was demonstrated by M\"uller\citep{m86}, who used a small network of
13 nuclei, from
helium to nickel. He showed how the inclusion of the temperature effects
was decisive in
avoiding the numerical oscillations which otherwise appear during and after
relaxation to the NSE regime. However, in that paper there were presented 
only very few test calculations, limited to isochoric combustion. Moreover,
the use of an entropy equation instead of the most frequently used 
energy equation was preferred to calculate the temperature evolution. 

In this paper we extend M\"uller's previous work to consider a larger nuclear
network and to analyze in greater detail the transition between time-dependent 
combustion
to complete nuclear equilibrium and viceversa. We also propose a method for
integrating based on the energy equation that is robust and straightforward to
implement, giving very 
good results in all the realized tests. In addition we have exhaustively 
checked our numerical scheme against a large variety of thermodynamic 
conditions of astrophysical interest. These include not only isochoric
combustion, as considered in M\"uller seminal work, bul also adiabatic, isothermal and isobaric processes, as well as the particular case of combustion 
within a strongly coupled coulombic plasma.

Nowadays
the effort to write the proposed numerical scheme for larger networks (even
of `kilo-nuclide' size) is considerably less than a decade ago, as we have the
great advantage of a recently published library of theoretical
reactions rates (i.e. Rauscher \& Thielemann~2000 or the
NACRE compilation, Angulo et al. 1999). An
efficient algorithm of this kind could be suitable for handling a variety of
vigorous nuclear combustion modes such as detonation, flames or silicon
photodisintegration. It is especially useful whenever the operator splitting
technique has to be applied to save computational time. Currently, this often
happens in
hydrodynamic calculations of both kinds of supernovae; in these situations the
implicit
thermal coupling could help to enlarge the time-steps without degrading 
neither the accuracy nor the stability of the system. There is also the
additional advantage to this scheme that it is more compatible with
the philosophy behind the post-processing technique. A necessary condition
for obtaining a detailed nucleosynthesis from the thermal history of an
explosive event, calculated beforehand using a reduced network, is that the
small and large networks give a similar energy release. Integrating the
chemical equations along with the energy conservation helps to fulfill this
requirement.  

In \S\ 2 we describe the method of calculation and its implementation in
two nuclear networks with 14 nuclei and 86 nuclei respectively. The
ability of the scheme to monitor the quasi (QSE) and complete statistical
equilibria is discussed. In \S\ 3 the results of three tests are provided:
adiabatic expansion from the NSE stage, hydrostatic and explosive silicon
burning and
nuclear flame propagation, all of them
of evident interest to current studies on supernovae. The role   
 of Coulomb corrections on the chemical composition of burned matter at 
 high densities is also discussed in \S\ 3.
Finally, \S\ 4 
is devoted
to summarizing the conclusions.

\section{Method used for integrating the nuclear network}

To integrate the set of differential equations associated with the nuclear
species we have chosen a scheme based on the simple and well checked implicit
method by
Arnett \& Truran \citep{at69}. The main motivation behind such election is
that 
Arnett\&Truran method has been traditionally used in nucleosynthetic
calculations, always giving satisfactory results with low computational effort.
In addition it is straightforward to 
modify in order to incorporate the thermal feedback and other proposed changes 
into the basic scheme. There are some differences between our integration 
method and that used in M\"uller\citep{m86} worth to mention. In M\"uller's
work the evolution 
of the system was followed by coupling the nuclear evolution with the entropy 
changes at constant density. Moreover, it was assumed that entropy changes
were only due to the release of nuclear energy, thus no other sources of
entropy were considered. Instead of the entropy equation our formulation uses 
energy conservation which is 
more often found in current calculations of explosive combustion.  
Moreover, the proposed 
algorithm is quite general, not restricted to isochoric processes, and  
potential entropy variations 
coming from heat exchange were also included. Minor effects, such as the 
incidence of coulombic corrections to the ionic component of the plasma and 
screening enhancement factors have been also incorporated to our scheme. A 
simplified description of our method can be found in Garc\'\i a-Senz\&Cabez\'on 
(2003).

 Starting from a typical differential equation governing the evolution of 
the molar fractions of the i-specie, $Y_i$:

\begin{equation}
\frac{dY_i}{dt}=\sum_{k,l}r_{kl}Y_k Y_l-\sum_j r_{ij} Y_i Y_j+
   \sum_m \lambda_m Y_m-\lambda_i Y_i
\end{equation}

\noindent
where $r_{ij}=\rho\mathrm{N_A}<\sigma,v>_{ij}$~stands for particle reactions and 
$\lambda_i$~ is for photodisintegrations and decays. 
As usual, reaction rates between identical particles have to be divided by the 
corresponding factor (two or six for binary and ternary reactions 
respectively). 
The triple-alpha reaction was also incorporated to equations (1). Other
possible three body reactions can be included in the same way. 
This set of equations is then linearized by taking

\begin{equation}
Y_{i}^{n+\theta}=Y_{i}^n+\theta\Delta Y_i
\end{equation}

\noindent
where $Y_i$\ is the molar fraction of the i-nuclei, n the integration step 
and $0\le\theta\le 1$~ a
parameter which allows the quality of the integration to be controlled.
Choosing
$\theta=1$~leads to a totally implicit first order integration;
$\theta=1/2$~is
centered (second order) implicit and $\theta=0$~is totally explicit. Then,
neglecting the terms
in $\Delta Y_i\cdot\Delta Y_j$,  the linearized version
of the chemical equations is written. With the same level of accuracy
we  propose to take $T^{n+\theta}=T^n+\theta\Delta T$~
and approximate the
nuclear reaction rates by taking
\begin{equation}
r_{ij}(T^{n+\theta})\simeq r_{ij}(T^n)+\frac{d r_{ij}}{dT}\Big]_{T^n}~
    \theta\Delta T
\end{equation}

\noindent
If we also neglect second order terms in temperature, such as 
$\Delta Y_i\Delta T$, we can approximate a typical element on the right
part of chemical
equations (1) by:  

\begin{eqnarray}
\nonumber
Y_{i}^{n+1}~Y_{j}^{n+1}~r_{ij}(T^{n+1})\simeq
\left[Y_i^n~r_{ij}(T^n)\right]\theta\Delta Y_j+
\left[Y_j^n~r_{ij}(T^n)\right]\theta\Delta Y_i+\\
\left[Y_i^n~Y_j^n
\frac{dr_{ij}}{dT}\right]\theta\Delta T+Y_i^n Y_j^n r_{ij}(T^n)
\end{eqnarray}

\noindent
then, equation (1) reads:

\begin{eqnarray}
\nonumber
\frac{\Delta Y_i}{\Delta t}=
 +\sum_{k,l}r_{kl}Y_l^n\theta\Delta Y_k+\sum_{k,l}r_{kl}Y_k^n\theta\Delta Y_l- 
    \left(\sum_j r_{ij}Y_j^n\right)\theta\Delta Y_i-\sum_j r_{ij}Y_i^n\theta\Delta Y_j+\\ \nonumber
    \sum_m\lambda_m \theta\Delta Y_m-\lambda_i\theta\Delta Y_i+
    \left(\sum_{k,l}r_{kl}^{'}Y_k^n Y_l^n-\sum_j r_{ij}^{'}Y_i^n Y_j^n+
    \sum_m\lambda_m^{'}Y_m^n-\lambda_i^{'}Y_i^n\right)\theta\Delta T+\\
    \sum_{k,l}r_{kl}Y_k^n Y_l^n-\sum_j r_{ij} Y_i^n Y_j^n+\sum_m\lambda_m Y_m^n-
    \lambda_i Y_i^n\qquad
\end{eqnarray}

\noindent
Note that the derivative of nuclear rates with respect temperature,
$r_{ij}^{'}$~and 
$\lambda_i^{'}$, do not pose a practical problem since
the recently published fitting
formulae of Rauscher \& Thielemann \citep{rt00} can be used to calculate
them in an efficient way. Using that compilation through all stages of the 
combustion also leads to a complete
internal
coherence between direct and inverse rates when the equilibrium is approached, which is an essential condition
for the proper handling of NSE.

For a network of $\mathrm N$ species we have $\mathrm N$ linear equations
with $\mathrm {N+1}$~unknowns. An
additional equation,  energy equation, is needed in order to close the
system. For a process not involving particle difussion it writes:

\begin{equation}
dQ=-P\frac{d\rho}{\rho^2}+\frac{\partial U}{\partial\rho} d\rho+
\frac{\partial U}{\partial T}dT+\sum_i\left(-\mathrm{BE}_i+
\frac{\partial U}{\partial Y_i}\right)dY_i  
\end{equation}

\noindent
where P is pressure, U is the energy per gram of material, 
$\mathrm {BE}_i$~represents the nuclear binding energy of a mol of 
the i-species and $dQ$~can also include sources or sinks of energy other than 
nuclear.

It is useful to write that equation by using the thermodynamical identity:

\begin{equation}
P=T\left(\frac{\partial P}{\partial T}\right)_\rho+
\rho^2\left(\frac{\partial U}{\partial \rho}\right)_T
\end{equation}

\noindent
thus, using $T^{n+\theta}=T^n+\theta\Delta T$, the  
discretized form of the energy equation is easily written. For example, for an
adiabatic process equation (6) reads:

\begin{eqnarray}
\sum_i\left(\mathrm{BE}_i-\frac{\partial U}{\partial Y_i}\right) \Delta Y_i-\left\{\left(\frac{\partial U}{\partial T}\right
)^n-
\frac{\Delta\rho}{\rho^2}
 \left(\frac{\partial P}{\partial T}\right)^n\theta\right\}\Delta T=  
 -T^n\frac{\Delta\rho}{\rho^2}
 \left(\frac{\partial P}{\partial T}\right)^n\qquad
\end{eqnarray}

\noindent
for an ideal gas of ions $\partial U/\partial Y_i=(3/2) \mathrm{N_AkT}$, which is usually 
much lower than the Q-value of the reactions. Coulomb corrections to the  
ideal gas must be also included when the ionic coupling constant
parameter, $\Gamma_i=2.27\times 10^5<Z^{5/3}>(\rho~Y_e)^{1/3}~T^{-1}$, becomes non 
negligible. This often happens in high temperature combustion of degenerate 
matter because of the 
strong dependence of $\Gamma_i$~on the mean charge of the system
. Such dependence also favours $\partial U/\partial Y_i$~of the products of 
charged particle reactions against that of the reactant particles. Thus, on the whole, 
both contributions to $\partial U/\partial Y_i$~(ideal gas and Coulomb 
corrections) work in the same direction,  
slightly increasing the equilibrium temperature for a given density. Therefore, regardless the density regime we have {\sl always kept}~ both contributions to $\partial U/\partial Y_i$~in equation (8).

Interestingly the implicit treatment of temperature also allows for the proper 
handling of quasi-hydrostatic and isobaric processes. Let's suppose that the 
value of the actual pressure of the system, P, remains always close to the 
external pressure, P$_\mathrm{ext}(t)$, so that $P=P_{ext}+dP$. Then, writing  
$dP=(\partial P/\partial T)dT+(\partial P/\partial\rho)d\rho+\sum_i
(\partial P/\partial Y_i)dY_i$~and taken out $d\rho$~the 
energy equation (eq.[6]) 
reads:

\begin{eqnarray}
\nonumber
dQ=\sum_i\left\{-\mathrm{BE}_i+\left(\frac{\partial U}{\partial Y_i}\right)
+\frac{T}{\rho^2} 
\left(\frac{\partial P}{\partial T}\right)
\left(\frac{\partial P}{\partial Y_i}\right)\left(\frac{\partial P}{\partial\rho}\right)^{-1}\right\}dY_i+\qquad\qquad\\ 
\qquad\qquad\left\{\left(\frac{\partial U}
{\partial T}\right)+\frac{T}{\rho^2}\left(\frac{\partial P}{\partial T}\right)^2
\left(\frac{\partial P}{\partial\rho}\right)^{-1}\right\}dT-(P-P_{\mathrm ext})\frac{T}{\rho^2}\left(\frac{\partial P}{\partial T}\right)\left(\frac{\partial P}{\partial\rho}\right)^{-1}
\end{eqnarray}

Once this equation is discretized (in a similar way that we did for eq.[6]) 
and solved
jointly with equations (5) the solution,  
$\Delta T$~and ${\Delta Y_i}$, can be used along with P$_{\mathrm ext}$~ to 
find $\Delta\rho$. Equation (9) has been used in \S 3.3 to solve the structure
of a nuclear flame. 

From here on we will focus on the case of
$\theta=1$~ which is expected to give the most stable system behaviour 
of the nuclear system. A recalculation of the first test presented in the 
next section with $\theta=0.5$~ did not lead to relevant quantitative changes in the results. The standard criterion to assign the integration time-step was 
to demand that the fractional change in temperature and molar fractions of 
the more abundant species would not exceed 2\% per model. Nevertheless, one can 
safely relax that criterion even to 10-15\% variation in molar fractions 
without seriously modify the
quantitative results. 

This set of linear equations can be efficiently solved by using standard
sparse matrix algorithms. We in particular used the one found in Prantzos, 
Arnould \& Arcoragi \citep{paa87}. The resulting algorithm led to a satisfactory mass
conservation, the error in the sum of mass fractions was always lower than $10^{-10}$, close machine precision, for the calculations described in
\S 3.
There is not a large loss of computational
efficiency when the linearized thermal coupling is turned on. The main source
of degradation of that type is the computation for the derivatives of the rates
rather
than the increase in the size of the matrix from $\{\mathrm N\times\mathrm N\}$~
to $\{(\mathrm {N}+1)\times(\mathrm{N}+1)\}$~. From the tests carried out
we estimate that our numerical
treatment leads to
a time overload under 15\% and 2\% for the
14 and 86-nucleus systems respectively.
Several tests were run which demonstrated the ability of
the scheme to monitor rapid isochoric combustion followed by adiabatic expansion
in the hydrodynamic time scale. Adding either diffusive terms or those
related to the artificial
viscosity formulation to the right side
of equations (6) and (9) allows us to include heat transport and shock wave heating
in the scheme. An extreme
case of heat transport, the propagation of a nuclear flame, is also
presented and
checked in \S\ 3. Even though all the test presented below were calculated 
solving the linearized equations (5) and (8) an improvement of the 
algorithm based on the iterative refinement of the solution is also given 
and discussed in \S 3.1.2.




\section{Testing the algorithm}

Three test cases were calculated in order to
check the method. All of
them are of an indisputable interest in explosive combustion: a) isochoric
combustion followed by adiabatic expansion once complete NSE has been achieved,
 b) endoenergetic
disintegration of silicon and
c) isobaric nuclear
flame propagation.

The nuclear network consists of 86 nuclei spanning from $^{12}\mathrm{C}$~to
$^{60}\mathrm {Zn}$~including
$\alpha$, p and n particles, as depicted in Figure 1. Even though this is not a
large network, but rather a moderate one, it is certainly  more
complex than the
$\alpha$-chain considered in M\"uller \citep{m86}. On another note, to
avoid an excessive computational burden, current
multidimensional simulations of supernovae can only support very reduced
networks. For this reason,  we also
carried out several runs of the test using an $\alpha$-network of 14
nuclei, a
subset of the previous sample of 86 nuclei. This reduced set was similar
to the sets  of 7 and 13 nuclei considered in the comparative study about the 
performance of small nuclear networks carried out by
Timmes et al. \citep{thw00}
. Nuclear rates on light elements
,up to $^{20}\mathrm {Ne}$, were taken from Caughlan \& Fowler \citep{cf88},
and from Rauscher \& Thielemann \citep{rt00} above neon.

A common feature of all the tests shown here is that nuclear burning proceeds
through the so-called e-process, thus it manages to achieve complete
nuclear equilibrium at high temperature. In nature, that equilibrium does not
last very long
because it is cut by the reduction in density as the expansion progresses.
At some point the temperature becomes too low to allow the nuclear reactions
to proceed and the chemical abundances freeze out. The  precise value of the
resulting abundances is a function of the initial entropy of the material in
the NSE
phase, the degree of neutronization of the material and the rate of the
expansion, as many studies have demonstrated (see for instance Meyer, Krishnan \& Clayton~
1998
for a recent 
work and references therein).

The equation of state (EOS) used in all the tests was quite complete, consisting
of a mix of electrons treated as a partially degenerate relativistic gas,
an ideal gas of ions with Coulomb and other minor corrections,  and
radiation. Although important at high densities, electron capture on protons
and nuclei have not been included in these
calculations. Nevertheless, they can be incorporated into the
scheme with the same formulation expressed in equations (2) and (3).

The treatment of electrostatic corrections deserves further discussion. At
high densities the energy of the plasma deviates 
from that of a pure ideal gas of nuclei and  electrons. Electrostatic
interactions must be taken into account in this regime, 
even in NSE, owing to the strong dependence of such interactions with the 
average charge of the system. The electrostatic energy of the k-specie 
associated to the
Coulomb interactions, $U_k^C$~(erg.~g$^{-1}$),  
can be approximated by the following expressions (Ogata \& Ichimaru~1987,
Yakolev \& Shalybkov~1987):

\begin{equation}
\frac{U_k^C}{\mathrm {N_AkT}Y_k}=
\cases
{a\Gamma_k+b\Gamma_k^{\frac{1}{4}}+c\Gamma_k^{-\frac{1}{4}}+
     d\qquad\Gamma_k\ge 1 \cr
-\frac{\sqrt{3}}{2}\Gamma_k^{\frac{3}{2}}+
  \beta\Gamma_k^{\gamma}\qquad
  \qquad\Gamma_k< 1 \cr}
\end{equation}

\noindent
where $\Gamma_k$~is the ionic coupling constant parameter of the k-nuclei, 
a=-0.8980, b=0.96786, c=0.2207, d=-0.86097, $\beta=0.29561$~and  
$\gamma=1.9885$. The 
chemical potentials, $\mu_k^C$, can be easily derived from expression (10). 
Nevertheless, the 
inclusion of Coulombic interactions affects not only the     
thermodynamics,  
they also slightly alters the rates of electron captures and the rate 
of nuclear reactions through the corresponding screening factors. As we have 
not included weak interactions in the network we will not address the first 
point. 
The second point, the incorporation of screening corrections to the nuclear rates in the scheme should be 
straightforward. However, there is also necessary  to compute the
derivatives of such corrections with respect to temperature, as we did with the 
nuclear reaction rates.  

In the following, we are mainly interested in the behaviour of the system when
it approaches the NSE. In this case it is enough to make a simplified treatment
of the screening. In 
particular, we take the enhancement factors given by 
Mochkovitch \& Nomoto \citep{mn86} who, among other things, studied the equilibrium 
of direct and reverse reactions when strong screening corrections are taken
into 
account. They found that the dominant term in the enhancement factor, EF, 
to direct capture reactions is:

\begin{equation}
\mathrm {EF}=\exp\left[\frac{\Delta\mu^C}{kT}\right]
\end{equation}

\noindent
where $\Delta\mu^c$~is the chemical potential of the 
reactants minus that of the products. The above expression is especially 
adequate for non-resonant reactions at high temperature as well as for resonant 
reactions with large enough resonance energy. Note that, in this approximation 
the use of the enhancement 
factors given by equation (11) does not affect to the photodisintegration
reactions calculated through the detailed balance, because 
the Q-value of the reaction  becomes  $Q\rightarrow Q_0+\Delta\mu^C$~which  
exactly 
compensates the screening correction. In the limit of complete nuclear 
equilibrium this approach leads to the same results as the statistical 
description given by the Saha equation with the chemical potentials corrected 
for electrostactic interactions (Bravo\& Garc\'\i a-Senz ~1999).

\subsection{Isochoric combustion and further expansion} 

Isochoric combustion is a very interesting test because the original
fuel (usually helium or carbon and oxygen in many explosive events) goes
through several
modes of burning: initial fast burning followed by  quasi-statistical
and
complete nuclear equilibrium. Further expansion will take the combustion
out of NSE, which leads to a rapid freezing of the abundances. In
hydrodynamic calculations, that regime might be associated to
a detonation wave passing through degenerate material.

To begin,  the
combustion of a system composed of equal parts $^{12}\mathrm{C}$~and
$^{16}\mathrm{O}$~at
$\rho=10^9$~g.cm$^{-3}$~was followed {\sl without} including the implicit
coupling
between abundances and temperature. The result is shown in Figure 2, where
we can see that mass fractions begin to {\sl oscillate}
as soon
as the combustion approaches the equilibrium regime. This behavior is 
characteristic of any stiff system of differential equations solved with explicit rather 
than implicit schemes,  
demanding  very short time-steps to
control the system ($<10^{-9}$~s). 
The evolution {\sl with}~thermal coupling is depicted in Figure 3; here the
oscillations {\sl did not show up} and the QSE and NSE regimes were handled
without
problems, despite the large time-steps ($\simeq 0.1$~s, even larger than the 
hydrodynamic timescale $\tau_{HD}$, see below)
taken once
equilibrium was
achieved. From Tables 1 and 2 we can see that the most abundant nuclei
correspond to
Fe-peak
elements, especially $^{54}\mathrm{Fe}$~(27\%, in mass) and also to helium
(almost 20\%). Such
abundances and the high temperature achieved during the isochoric combustion are
in agreement with the current picture of NSE (Arnett~1996).

The integration scheme also allowed us to follow a rapid adiabatic expansion in
the hydrodynamic timescale $\tau_{HD}=446/\sqrt\rho$~s immediately after
the NSE regime was reached:

\begin{equation}
\rho(t)=\rho_0~\exp[-t/\tau_{HD}]
\end{equation}

During the expansion there was no need to impose
any particular freeze-out temperature for the nucleonic system. 
In Table 1 we provide the mass fractions at five different
fiducial temperatures achieved along the cooling path. From these numbers we 
can see that there is a large spread in the freezing temperatures. Many 
species are still changing even below $T=2\times 10^9$~K, although their abundances 
are generally too low to be significative. The most abundant element, 
$^{56}\mathrm{Ni}$, freezed around $T\simeq 3\times 10^9$~K which is not a high 
enough temperature to keep the NSE going. Thus, whenever the 
Saha equation is used  
 to describe nuclear statistical equilibrium above $4-5\times 10^9$~K one has to be
cautious and switch to a time-dependent network calculation below that 
critical temperature because the material is not completely burned yet. 

In Table 2 we provide the freeze-out temperature and final composition for
each of the five most abundant nuclei as a function of the initial
thermodynamical 
 state and chemical
composition. As one can see, while the critical temperature for achieving
complete NSE was the same for all nuclei this was certainly not true for the
freeze-out temperature. When the initial fuel was composed of carbon and oxygen
with no neutron excess, $\eta=0$~($\eta=1-2Y_e$), the resulting abundances
were largely dominated by $^{56}\mathrm{Ni}$. Under the conditions of our 
numerical 
experiment, $^{56}\mathrm{Ni}$~froze at $T_9=3.28$, while other Fe-peak isotopes  
were still reacting with residual protons and neutrons. Therefore their freezing
temperatures were lower. A similar calculation (same initial density and temperature)
is shown in the second row of Table 2, but the initial
composition was $30\%$~$^{12}\mathrm{C}$, $30\%$~$^{16}\mathrm{O}$~and
$40\%$~$^{22}\mathrm{Ne}$, resulting in  
a moderate neutron excess of $\eta=0.0364$. This time the dominant nucleus
after
the adiabatic cooling was not $^{56}\mathrm{Ni}$~but instead were
$^{54}\mathrm{Fe}$~and
$^{58}\mathrm{Ni}$~in
accordance with well-known results (i.e., Clayton~1968). There is, however,
a clear
difference in the freezing temperatures of these dominant isotopes.

The third row of Table 2 and Figure 4 summarizes the evolution of a system
initially composed of pure helium at $\rho=4\times 10^6$~g.cm$^{-3}$. In this  
case $\eta=0$~and the combined value of temperature and density in
NSE resulted in a larger entropy than in the two earlier cases. 
Now, the period of relaxation to NSE was larger than in the two 
precedent tests; however there was no sign of instability during the 
integration. Once the NSE was achieved, at $T=4.3\times 10^9$~K, the composition
 was again 
dominated by Fe-peak elements with a lower concentration of light particles 
$\alpha$~and p, (0.66\% and 0.74\% by mass respectively) 
than in the two precedent cases. For a given initial density the outcome of
the adiabatic cooling depends on the adopted expansion rate. For very fast 
expansions the alpha particles have some difficulty to react, owing the 
low-density environment, and the final distribution is rich in $\alpha$~
particles and intermediate-mass elements
 (Woosley, Arnett \& Clayton~1973). In our calculation the expansion rate 
was not rapid enough to allow the synthesis of intermediate-mass elements
 in appreciable amounts. According to Table 2 the final composition was
 completely 
dominated by $^{56}\mathrm{Ni}$~(98\%) and Fe-peak elements plus a little 
amount of unburnt helium. 

\subsubsection{Influence of screening factors in the equilibrium distribution 
of nuclei}

As mentioned above, the enhancement factors to the nuclear reaction rates given
by equation (11) were taken into account in all the realized tests. Generally 
speaking, their inclusion in explosive processes, although important, is not
very crucial; its 
main effect being to shift the temporal evolution and the maximum 
value achieved by temperature. Therefore, there is expected to be some 
changes in the mass fractions of the species in NSE when screening factors are 
incorporated into the scheme, especially at high density. In Figure 5 there is 
represented the evolution of temperature during the isochoric combustion of 
a 50\%-50\% mix of carbon and oxygen at constant density of
$4\times 10^9$~g.cm$^{-3}$. As we can see the profiles with and without enhancement 
factors are not very 
different, although the induction time is shorter and the equilibrium
temperature is no longer the same:  
$T_9=9.51$~ (no screening included) and $T_9=9.77$~(screening included). Again, there 
is a stable numerical behaviour of the system when the equilibrium is
approached.

To analyze  the  
nucleosynthesis during the NSE stage it is more convenient to compare  
at the same temperature. In doing that not only the 
comparison among abundances becomes more reliable, but a direct check with the 
results obtained by using the nuclear Saha equation with the mass excesses 
corrected from the coulombic part of the chemical potentials deduced from equation (10) is also possible. The results of such comparison for the 14-nuclei 
network are depicted in Figure 6. From that figure it becomes clear that the 
main effect of screening corrections is to shift the distribution of the 
species towards higher atomic masses and that, the larger the density the 
bigger the change. In fact, the change in the NSE distribution of the nuclei  
depicted in Figure 6 looks rather spectacular. This is in part due to the
limited size of the 
14-nuclei chain; when the same calculation was repeated with the 86-nuclei 
network such differences were {\sl spread out and smoothed}. For example, 
at $T_9=9.32$~$, \rho_9=4$~the mean molecular weight calculated from the $\alpha$-network
was
$\mu_i=11.04$~g.mol$^{-1}$~(no 
screening factors) and 
$\mu_i=14.35$~g.mol$^{-1}$~(with screening factors) whereas the corresponding values for the 
86-nuclei network were $\mu_i=10.54$~g.mol$^{-1}$~ and $\mu_i=11.87$~g.mol$^{-1}$~ respectively. 
The comparison 
between the resulting mass fractions calculated through the network and the 
nuclear Saha equation also confirms that the adequate way to handle the 
coulombic interactions in the NSE regime  
is by substracting the electrostatic chemical potentials from the 
nuclear mass excesses (Bravo\&Garc\'\i a-Senz 1999).  

\subsubsection{Iterative refinement and accuracy}

The integration scheme developed in \S\ 2 can be considered as the first
iteration step of a more general Newton-Raphson based method for solving the
full non-linear equations governing the evolution of the nuclear system. 
When using these methods there is often an enhancement of the computational 
efficiency despite that the time invested per model is larger. In addition 
they also have the advantage that a check on accuracy is possible, allowing 
for a more consistent choice of the time-steps (Timmes 1999). Fortunately 
it is straightforward to modify our numerical scheme, with minimum changes,  
in order to allow the 
iterative refinement of the solution, as described below. The resulting 
algorithm is equivalent to the standard implicit multidimensional 
Newton-Raphson calculation. 
For a given $n+1$ time-step the iterative process leads to 
a sequence of 
`solutions' $(Y_i^{n+1})^k$~and $(T^{n+1})^k$~for the i-specie and 
temperature which, after $m$~ iterations, should converge to the true
values for $Y_i^{n+1}$~and
$T^{n+1}$. We define a set of functions $G_i^0$, which represents the 
discretized form of the chemical equations (1):  

\begin{eqnarray}
G_i^0[({\bf Y}^{n+1})^k,(T^{n+1})^k]=
F_i[({\bf Y}^{n+1})^k,(T^{n+1})^k]~\Delta t-
[(Y_i^{n+1})^k-Y_i^n]\qquad  k=1...m
\end{eqnarray}

\noindent
where $({\bf Y}^{n+1})^k=\{(Y_1^{n+1})^k....(Y_N^{n+1})^k$\}~and $F_i$~are the terms on the right of equation (1). A similar function   
,$G^1$, representing the energy equation (8) is also defined:  

\begin{eqnarray}
\nonumber
G^1[({\bf Y}^{n+1})^k,(T^{n+1})^k]=
\sum_i\left(\mathrm{BE}_i-\frac{\partial U}{\partial Y_i}\right)
   [(Y_i^{n+1})^k-Y_i^n)]\qquad\qquad\qquad\\ 
  \qquad\qquad-\left\{\left(\frac{\partial U}{\partial T}\right)-
\frac{\Delta\rho}{\rho^2}
 \left(\frac{\partial P}{\partial T}\right)\right\}
 [(T^{n+1})^k-T^n]+  
 T^n\frac{\Delta\rho}{\rho^2}
 \left(\frac{\partial P}{\partial T}\right)
\end{eqnarray}

The Newton-Raphson iterative method demands that 
$\delta G_i^{0,1}=-G_i^{0,1}$, which allows to solve for 
the 
corrections $\Delta Y_i, \Delta T$~to the approximate values $(Y_i^{n+1})^k$~
and $(T^{n+1})^k$~found in the latest iteration.   
In this case the Jacobian matrix 
associated to $\delta G_i^{0,1}$~is {\sl exactly the same} as the matrix of the
linear 
system associated to equations (5) and (8).
For the first
iteration we take $(Y_i^{n+1})^1=Y_i^n$~and $(T^{n+1})^1=T^n$~so that when m=1 
the extended Arnett\&Truran
method given by equations (5) and (8) is recovered.
For $m>1$~an iterative sequence results which usually ends when 
the $k$-corrections  $\Delta Y_i^k$~to $(Y_i^{n+1})^{k-1}$~and 
$\Delta T^k$~to $(T^{n+1})^{k-1}$~ becomes negligible or, better, when
the $G_i^{0,1}$~functions defined by equations (13) and (14), conveniently 
normalized, go to zero (for example we took  
$|G_i^0|/\sqrt{(F_i\Delta t)^2+\Delta Y_i^2}<5\times 10^{-6}$~and a similar expression for
$G^1$) . Usually  
convergence 
is achieved in 3-4 iterations. The criterion to choose the time-step was not 
very restrictive, we allowed for a $1.5\%$~relative variation in temperature 
and $10\%$~in molar fractions. In Figure 7 there is 
shown the evolution of the errors in mass fractions of several species 
during the isochoric carbon combustion, estimated by comparing the size   
of the corrections given by the first iteration with the algebraic sum of the 
lowest-order corrections given by the remaining iterations.  
As we can see in that figure the first iteration always led to 
a much larger correction, usually of two orders of magnitude or more, than that
of the sum of the remaining iterations. Thus, the evolution of the mass 
fractions and temperature can reasonably be described taking only one iteration and neglecting second and higher-order corrections.
For this particular test, a further relaxation of the criterion that restrict 
the time step 
led to several negative abundances. Nevertheless,    
the iterative Newton-Raphson scheme would also be in trouble in this case, due 
to the lack of convergence. Therefore, taking only one iteration along as a 
moderate time-step would generally be enough, especially  
if we are mainly interested in the thermal evolution of the system and 
in the gross features of the nucleosynthesis. For larger time-steps or  
for detailed nucleosynthetic studies it is better to rely 
on the iterative scheme which has the additional advantage that a check of the 
accuracy is feasible (Timmes 1999).

\subsection{Hydrostatic and explosive silicon burning}

Hydrostatic combustion of silicon is a quasi-equilibrium process from the
beginning due to the important role played by photodisintegration reactions.
One goal of this calculation is to check the ability of our integration
scheme to describe silicon photodissociation at constant temperature and
density. We also studied  explosive silicon burning within an
expanding medium, until  the
reactions were totally quenched. In both cases we compared our results
with those
of Timmes et al. \citep{thw00}, which were calculated under
the same
environmental
conditions. Thus, a system composed of 100\% $^{28}\mathrm {Si}$~at
T=$6\times 10^9$K and
$\rho=10^7$~g.cm$^{-3}$~was considered. In the
hydrostatic case, 
the temperature was kept constant, therefore the feedback between
nuclear kinetics and thermal evolution no longer took place. Nevertheless,
our scheme was also able to keep track of this limiting case without introducing
any
modification. Instead to rewrite our code to not include $\Delta T$~terms in
the matrix we have artificially raised the value of the specific heat during 
the silicon dissociation to keep the temperature constant. In  
the adiabatic expansion test the specific heat was again restored  to
its real value.

The results of these calculations for the small network are shown
in Figures 8,9 and 10.
A comparison between the evolution of the nuclear energy generation rate,
provided in Figure 8, with respect to that shown in Figure 1a of Timmes et al. \citep{thw00}
is satisfactory.
The main difference
is the tiny bump shown around t$\simeq 2\times 10^{-4}$~s, which was probably due
 to the
different libraries used to compute the nuclear rates. A comparison of their
abundances (their Fig. 1c) with ours (Fig. 9) also shows a good agreement.

As a second check a piece of  pure
silicon was enforced to follow an adiabatic expansion in the hydrodynamic
timescale as shown in equation (12). In this case, the
temperature evolution was calculated self-consistently from the network. The
results of this calculation for our $\alpha-$chain are summarized in Figures 
8 and 10. These figures can be compared with Figures 2a,2c in Timmes et al.
\citep{thw00}, in which
the temperature is evolved with an approximate analytical formula. The evolution of
the nuclear
energy rate follows a close path in both calculations. The comparison
between  abundances is also quite satisfactory, although there are
divergences above t=0.1 s, which is about the hydrodynamical time $\tau_{HD}$.
These differences probably arise from the
slightly different path followed by temperature in both calculations. We note 
that Timmes et al. (2000) only included the contribution of the radiation to
the specific heat thus, for 
a given density, we found a higher value of temperature.
In particular, our final  silicon
group abundance (up to titanium) was about five times higher than theirs. On
the whole its
composition is characteristic of an $\alpha-$rich freeze-out
with a large abundance of $^{56}\mathrm{Ni}$~(92\% in mass) followed by
$^4\mathrm{He}$~(7\%).

\subsection{An example of isobaric combustion: The propagation of a 
subsonic nuclear flame}

Thermonuclear flame propagation is a subsonic mode of
energy propagation which takes place under nearly isobaric conditions.
Flames could appear due to the combination of sharp thermal gradients,
high conductivity and potentially explosive fuels such as helium or carbon
and oxygen. These requirements may be fulfilled in the interior of those massive
 white
dwarfs that are reactivated by the accretion from a companion star, giving
rise to a Type Ia supernova.
This calculation is more complex than those carried out in the 
previous subsections because both,  the temporal evolution and the spatial 
structure of the flame have to be solved. Therefore our nuclear subroutine was 
coupled to the diffusion equation within an initially uniform grid of shells
representing a
physical 
system with planar geometry and constant density. 
Given the adequate initial conditions (see below), the advance
of the flame was calculated directly from the nuclear network of 86 isotopes
along
with the isobaricity assumption (Timmes \& Woosley~1992).
A diffusive term was
added to the energy equation so that the heat is transported from the
burned to the unburned zone, mainly by electronic conduction. Thus, we take,  

\begin{equation}
dQ=\dot\mathrm S_{dif}~dt=\frac{1}{\rho}{\bf\nabla}\cdot(\sigma
{\bf\nabla T})~dt
\end{equation}

\noindent
in equation (9) and set $P_{ext}=\mathrm {constant}$~in the same equation in
order to
keep the pressure constant during the flame propagation. The conductivity,  
$\sigma$, was that of Khokhlov, Oran and Wheeler \citep{kow97} which includes 
electronic and radiative contributions to the opacity.
As the
dependence of conductivity on temperature is much lower than
it is in  nuclear
reactions this new term, equation (15),  
can be added to the independent terms in the corresponding discretized 
version of equation (9) . In order to
build a nuclear 
flame the left side of a tube containing equal parts
of carbon and oxygen was incinerated at constant pressure until
NSE was achieved. After a brief transitory period the heat exchange between the
burned  the and cold material gave rise to a steady planar flame that moved
at a constant velocity through the fuel with constant density
$\rho_{fuel}=1.26\times 10^8$~g.cm$^{-3}$.
Behind the flame front the original fuel was rapidly transformed into
intermediate-mass elements, especially silicon; the temperature rose to
$6.1\times 
10^9$~K; and the density dropped to $\rho=7.6\times 10^7$~g.cm$^{-3}$~to keep
the
pressure constant. Under these conditions the ashes were in QSE and
began to slowly relax to complete equilibrium, as can be seen in Figure 11.
During the total elapsed time (about $10^{-5}$~s) the nuclear system did not
show any sign of instability. The stationary velocity of the flame front was
$\mathrm v_{flame}=4.8$~km.s$^{-1}$, which is about 50\% higher than that
obtained by Timmes \& Woosley \citep{tw92} for the same density. Such
discrepancy is well within the factor two error usually attributed to the 
input physics (conductivities, nuclear libraries and EOS). 

\section{Conclusions}

Current standard integration methods for chemical equations encounter some
difficulty when handling high temperature combustion near equilibrium. The
reason 
for this is the extreme sensitivity of nuclear rates to the temperature, which
makes it
difficult to reach a time-independent situation (equilibrium) when using a
time-dependent solver for equations that describe the evolution of molar
fractions. M\"uller \citep{m86} pointed out that the solution is to
implicitly couple
the equations that give abundance changes with those that describe
energy or entropy
evolution. A practical drawback of such a procedure is that the
derivatives of the nuclear reaction rates have to be worked out at each
iteration. 
Nowadays, however, very complete libraries are available, which detail
nuclear reaction rates for many species with simple formulae
(i.e. Rauscher
\& Thielemann~2000). From these, then,
derivatives can be evaluated without causing excessive computational
overload.

The main
purpose of this study was to build and check an integration method, inspired by
M\"uller idea, that would be
flexible enough to accommodate a variety of combustion modes
involving
quasi/total nuclear equilibrium at some point along the path of its evolution.
Screening enhancement factors were also incorporated to our numerical scheme. 
Good
astrophysical examples of the potential applicability of this algorithm would
be for
nuclear flames
and detonation waves in Type Ia supernovae and for silicon burning in Type II
supernovae respectively.

In spite of the simplicity of the proposed numerical algorithm, described in 
\S 2,
we found that it was able to pass many tests related
to high-temperature combustion, \S 3. In each of them stability was
always preserved, even in the NSE stage, without restricting the
time-step. The results also matched well those obtained in similar
conditions by other authors using
 different integration schemes. A comparison of the numerical efficiency
of the small chain (of 14 nuclei)
 and the moderate-sized one (of 86 nuclei) showed that the
time overload introduced by thermal coupling was under 15\%, and 2\% for
the small and large networks
respectively. These factors arise mainly from the calculation of the 
derivatives of the nuclear rates rather than from the inversion of the
matrix. Recently, the 14-isotope $\alpha$-chain with thermal coupling was
incorporated to a multidimensional SPH code designed to simulate Type
Ia supernova
explosions. In that situation, the average time degradation is far below the
aforementioned
15\% as the time expended per model is largely controlled by
hydrodynamics,
especially when one is searching for neighbours or calculating gravity.

The FORTRAN subroutines that implement the 14 $\alpha$-chain and
the 86-isotope network presented in this paper are made available by
the authors to
interested readers upon request.

\acknowledgments

This work was funded in part by MCYT grants EPIS98-1348 and AYA2000-1785;
and CIRIT GRQ grants.




\clearpage


\begin{figure}
\plotone{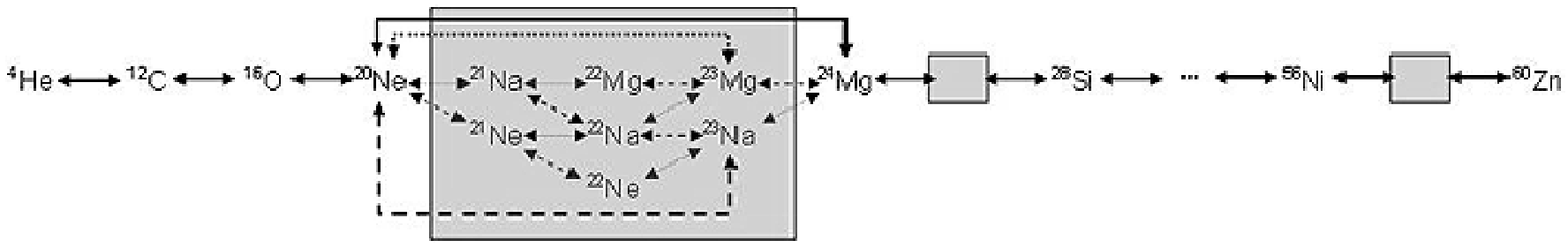}
\caption{Sketch of the nuclear network used in the tests. It consists of 86
nuclei (including p, n and $\alpha$~particles), linked trough (p,$\gamma$)
-thin solid arrows-, (n,$\gamma$) -small dotted arrows-, ($\alpha$,$\gamma$)
-thick solid arrows-, ($\alpha$,p) -long dashed arrows-, and ($\alpha$,n)
-long dotted arrows-, as suggested by the plot. The network is
organized into groups around an $\alpha$-chain of 14 nuclei from $^4\mathrm {He}
$~
to $^{60}\mathrm {Zn}$. One of these groups, between $^{20}\mathrm {Ne}$~and
$^{24}\mathrm {Mg}$, is shown in the figure. Triple alpha reaction and binary
reactions of carbon
and oxygen were also included.\label{fig1}}
\end{figure}

\clearpage 

\begin{figure}
\plotone{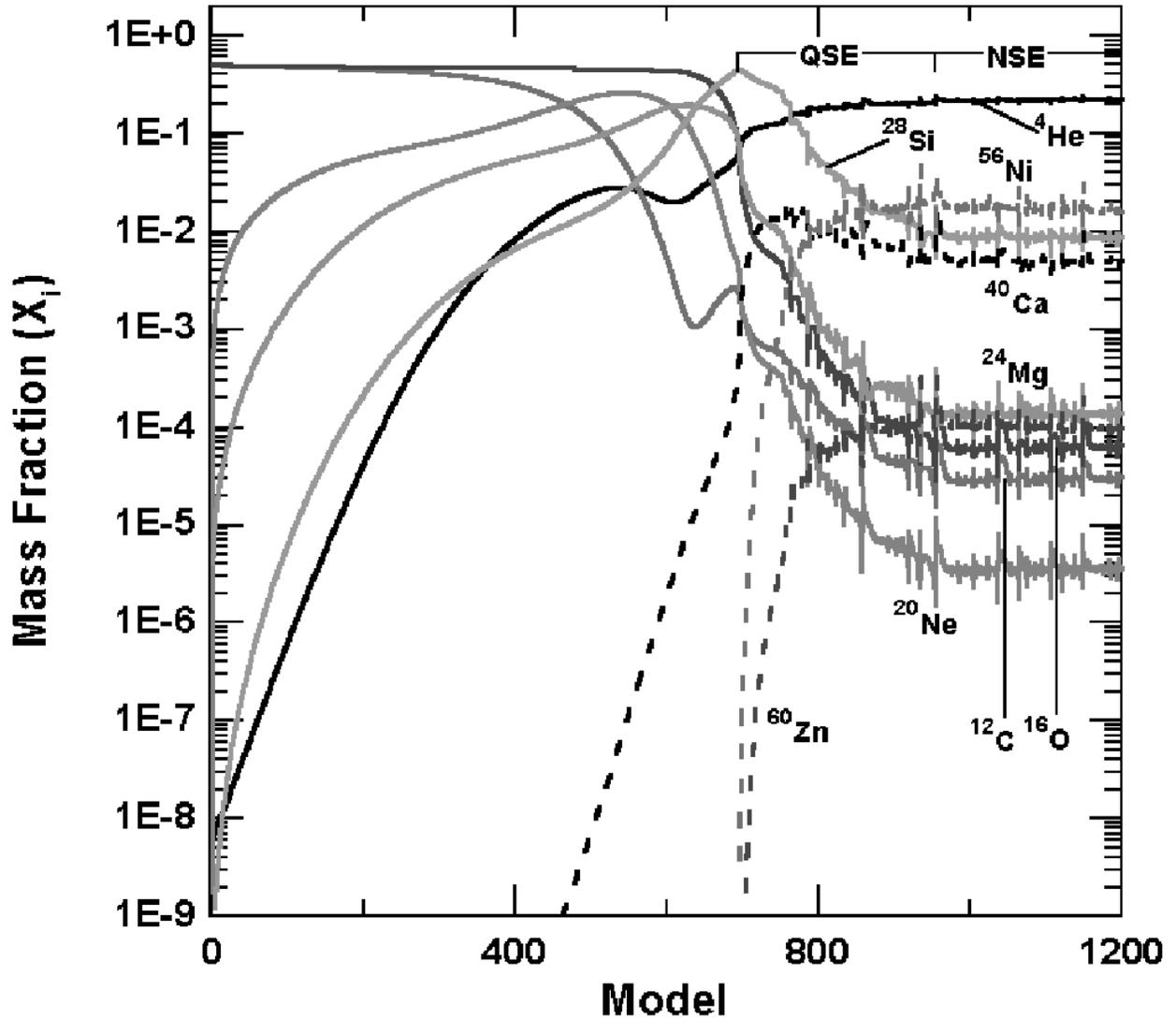}
\caption{Evolution at constant density, $\rho=10^9$~g.cm$^{-3}$~ 
of a 50\% mix
of carbon and oxygen calculated {\sl without}~implicit thermal coupling, the 
initial temperature was $T_0=10^9$~K. As
soon as the combustion enters in the QSE zone and NSE plateau it becomes
unstable
 (86-isotope network).
\label{fig2}}
\end{figure}

\clearpage

\begin{figure}
\plotone{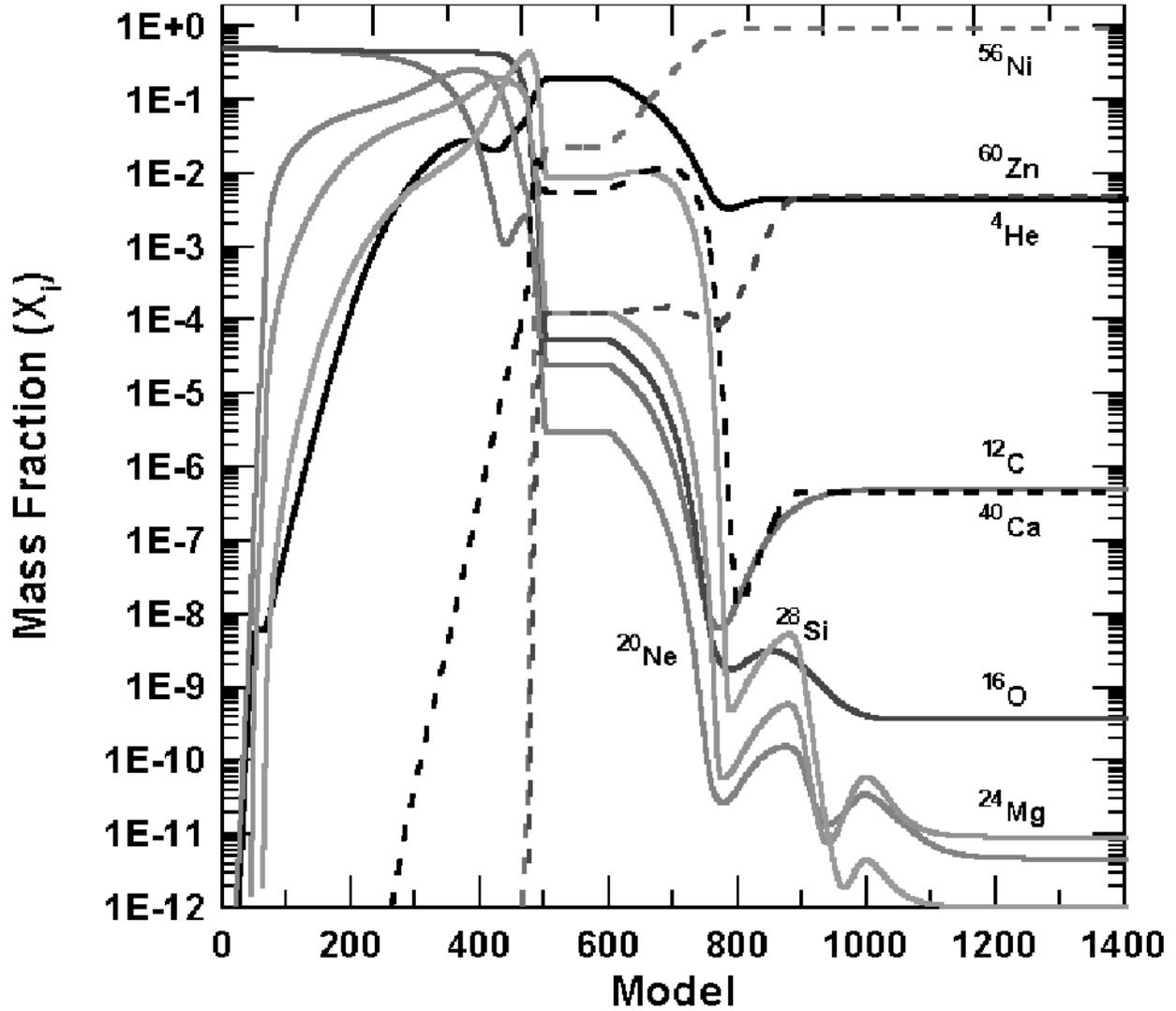}
\caption{The same as in Figure 2, this time with the thermal
coupling {\sl turned
on}. The system remains stable all the time. An adiabatic expansion was
imposed above model 600 until all the nuclear reactions quenched
(86-isotope network).
\label{fig3}}
 \end{figure}

\clearpage

\begin{figure}
\plotone{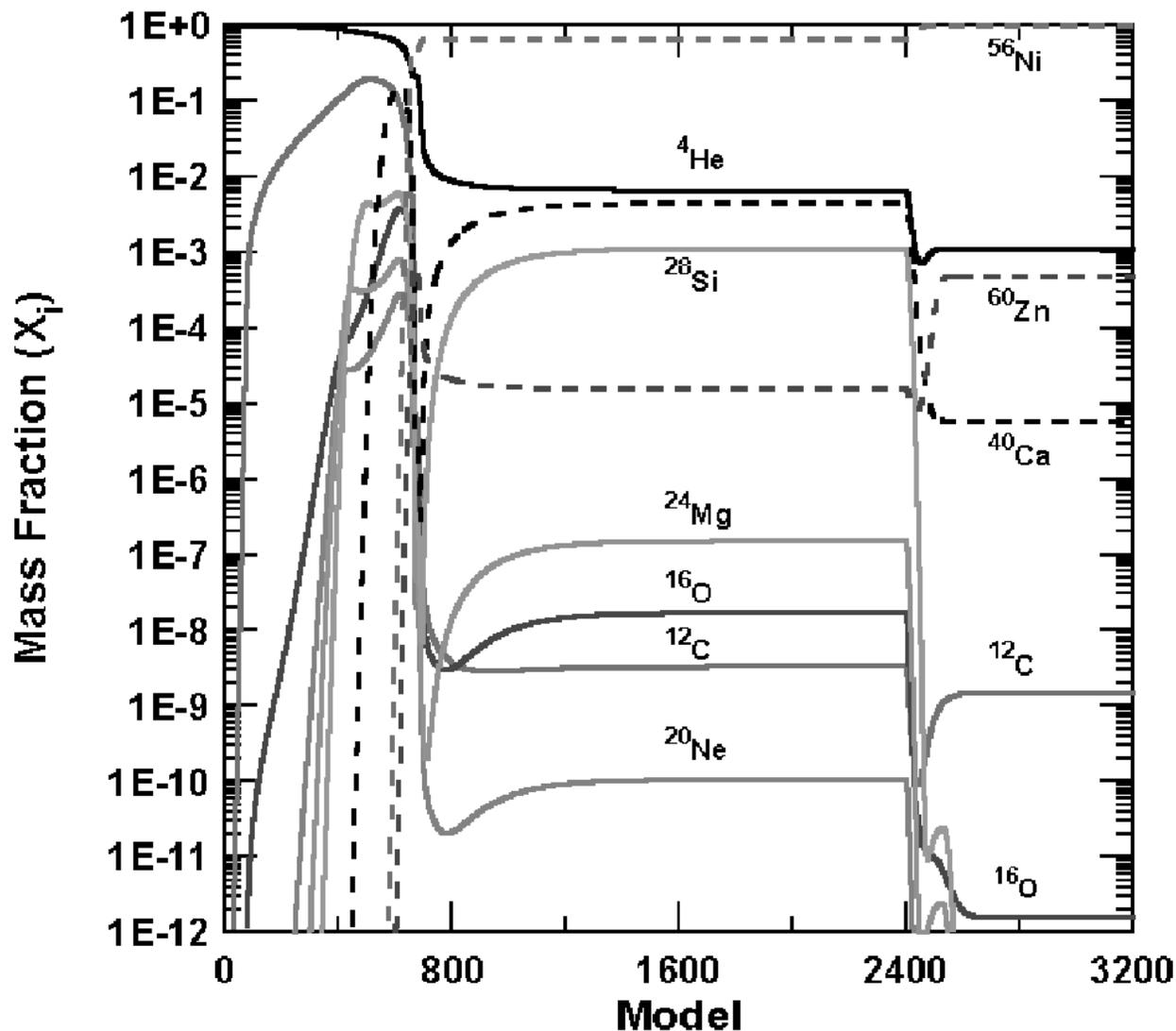}
\caption{Isochoric combustion of pure helium (at $\rho=4\times 10^6$~g.cm$^{-3}$)
followed by
an adiabatic expansion in the hydrodynamic time. Owing to the lower ignition
density it took more time for the combustion to reach complete equilibrium
at $\mathrm{T}=4.23\times 10^9$~K.
Nevertheless, the implicit thermal coupling again avoids any instability to
grow.
The resulting nucleosynthesis after the expansion is very rich in
$^{56}\mathrm{Ni}$, with appreciable amounts of helium and 
minor concentration of $\alpha$-elements such as
$^{40}\mathrm{Ca}$~(86-isotope network).
\label{fig4}}
\end{figure}

\clearpage

\begin{figure}
\plotone{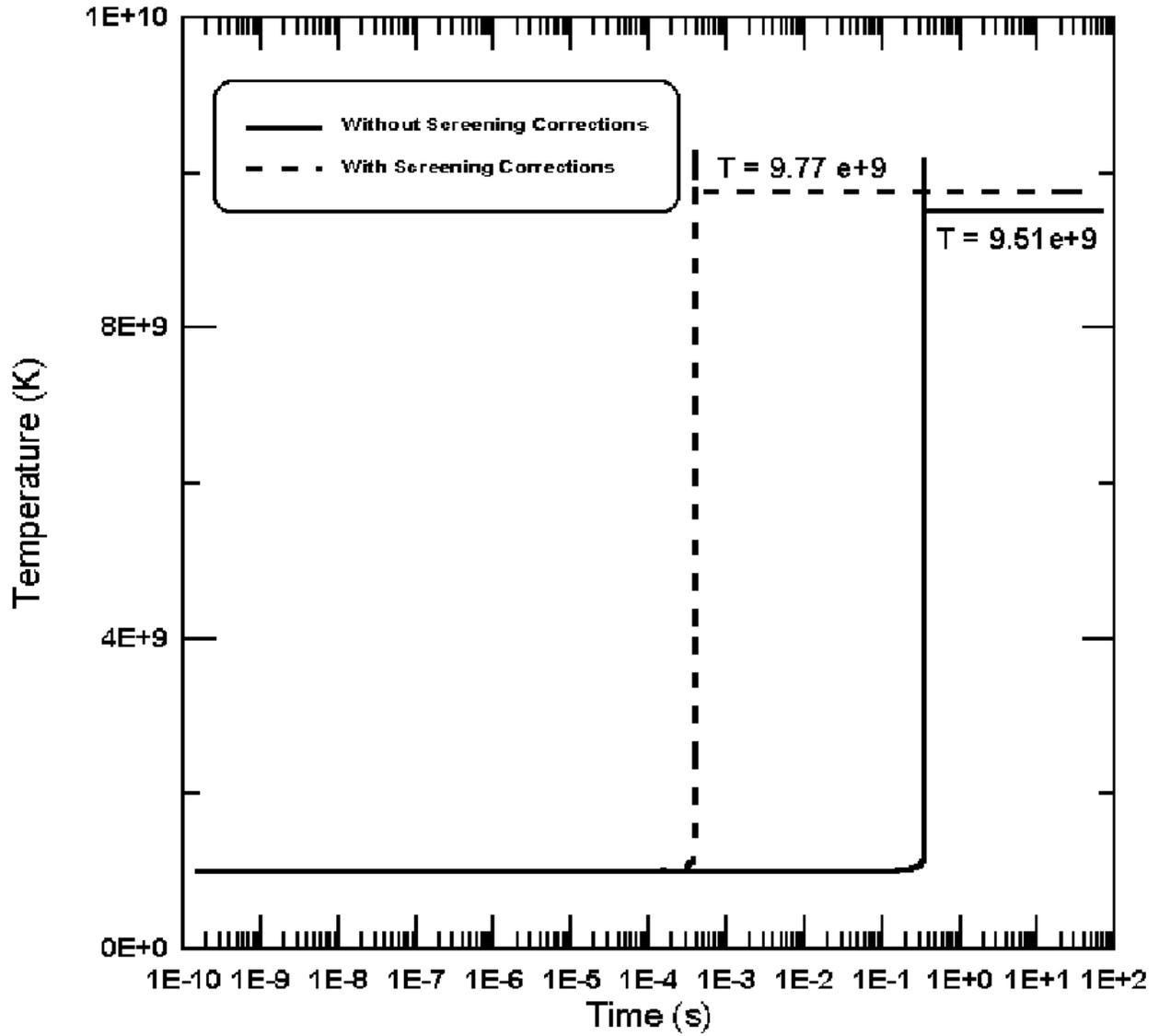}
\caption{Influence of the screening corrections on the evolution of temperature
during
the isochoric combustion of a 50\% mix 
of carbon and oxygen at $\rho=4\times 10^9$~g.cm$^{-3}$. Inclusion of screening 
factors (dashed line) led to a higher temperature value during the NSE.
 (14-isotope chain).
\label{fig5}}
 \end{figure}

\clearpage

\begin{figure}
\plotone{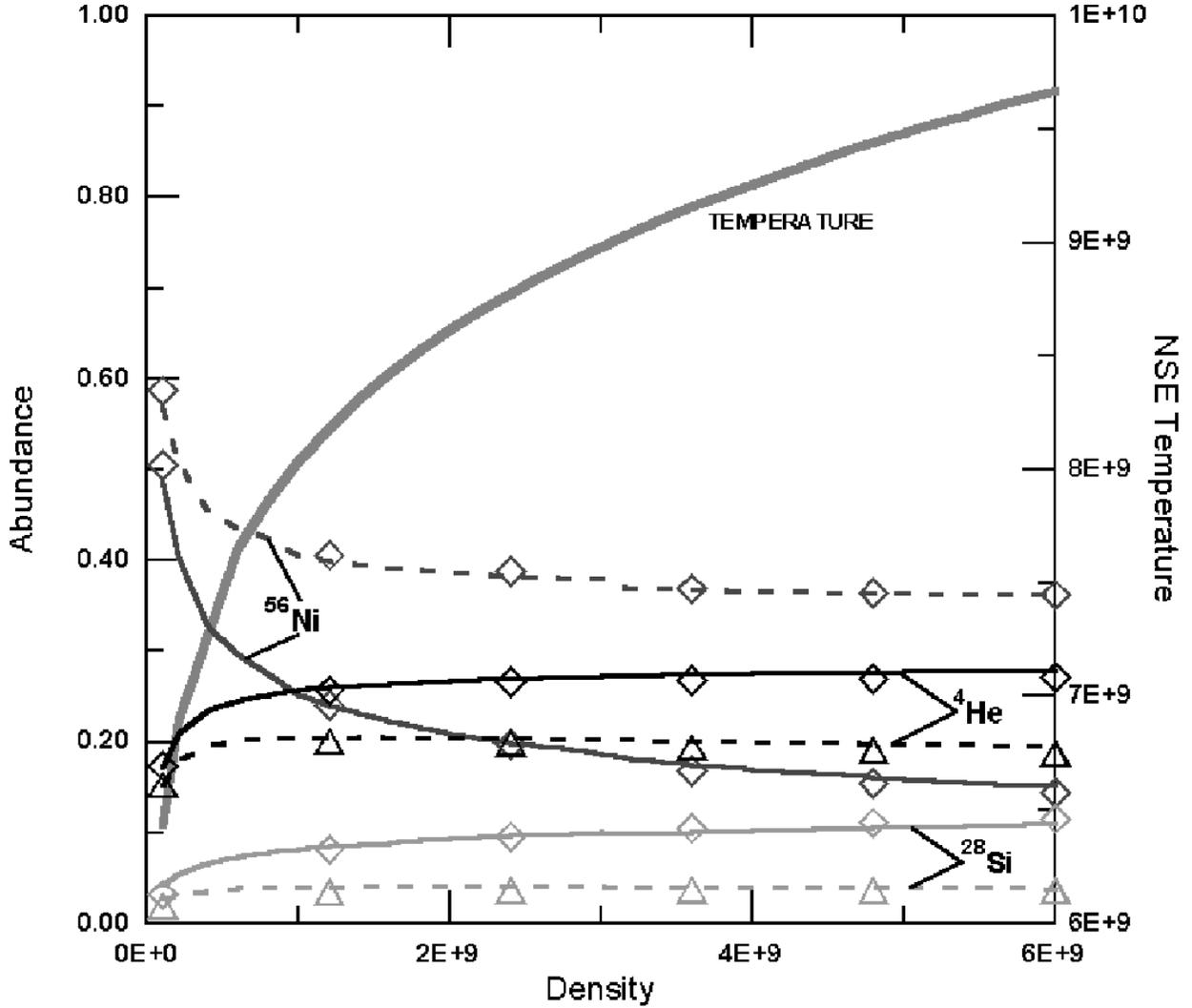}
\caption{Abundance of $^4\mathrm{He}$, $^{28}\mathrm{Si}$~and $^{56}\mathrm{Ni}$~at several densities 
once the NSE has been achieved. For a given density and temperature the 
isotopic composition at NSE sensitively depends on the inclusion of the 
screening corrections to 
the nuclear reaction rates. Continuum lines depicts the abundances when no 
screening factors are taken into account. Dashed lines are for the screening 
factors given by equation (11). A comparison with the mass fractions obtained 
by solving the nuclear Saha equation for the same number of species 
(14) is also provided (triangles and 
diamonds). In this case 
the electrostatic interactions were taken into account by substracting the 
chemical potential of the nuclei from their nuclear mass excess. There is a 
good agreement 
between both calculations. As expected, the effect of the electrostatic 
interactions increases with density. (14-isotope chain).
\label{fig6}}
 \end{figure}

\clearpage
 
\begin{figure}
\plotone{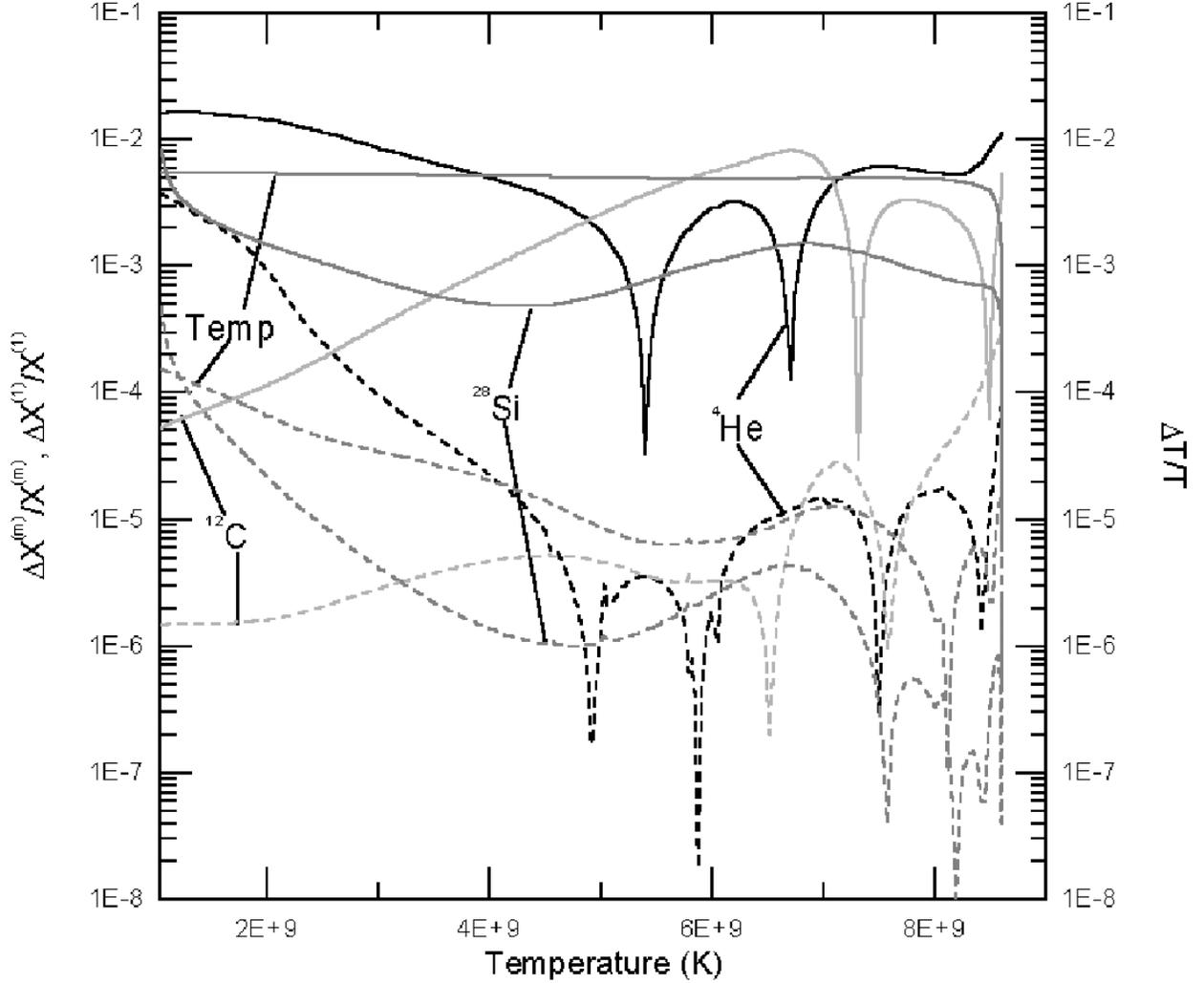}
\caption{Evolution of $\Delta X_i^1/X_i$~(solid lines), and 
$\sum_k\Delta X_i^k/X_i,~ k=2,m$~(dashed lines) corresponding to 
 He$^4$, 
C$^{12}$~and Si$^{28}$~and the equivalent expressions for temperature, 
during the 
isochoric combustion of a $50\%$~mix of carbon and oxygen at  
$\rho=10^9$~g.cm$^{-3}$. Horitzontal axis is temperature, which goes from 
$T^0=10^9$~K to $T=8.6\times 10^9$~just before the relaxation to NSE.
The error during the purely linear calculation (one iteration) is approximately given by the sum of the higher-order corrections (dashed lines) 
calculated through the 
iterative Newton-Raphson scheme described in \S 3.1.2.
The time-step was chosen to allow a relative variation of $1.5\%$~
and $10\%$~in temperature and mass-fractions of the more abundant species 
respectively (86-isotope chain).  
\label{fig7}}
 \end{figure}

\clearpage
\begin{figure}
\plotone{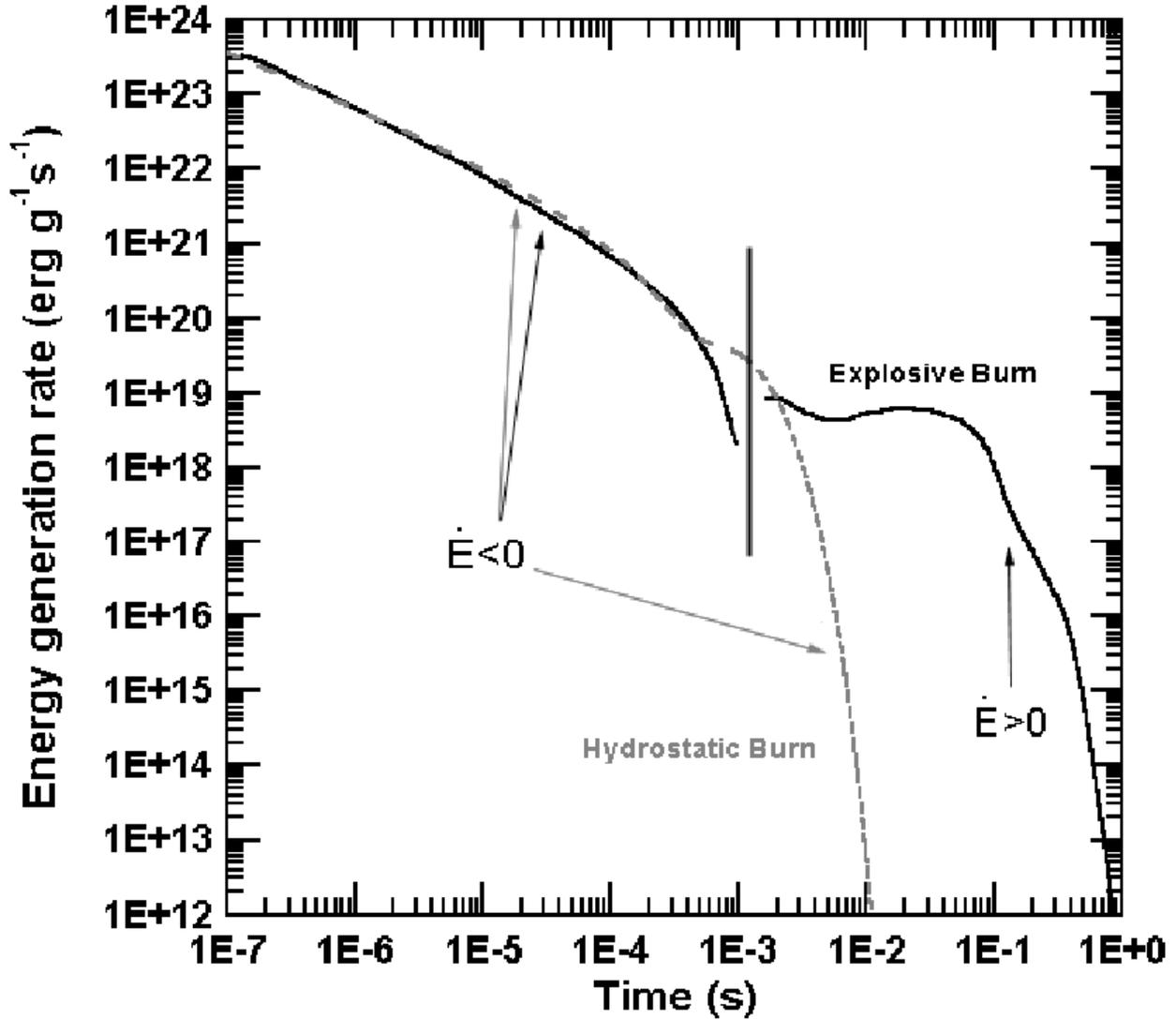}
\caption{Evolution of the nuclear energy generation rate (absolute value) during
endoenergetic silicon combustion:~ 
at constant density $\rho=10^7$~g.cm$^{-3}$~and temperature $\mathrm{T}=
6\times 10^9$~K (dashed line), and during the explosive combustion (solid line). For
the last case the nuclear energy generation rate becomes positive after
$10^{-3}$~s, (14-isotope chain).
\label{fig8}}
 \end{figure}

\clearpage

\begin{figure}
\plotone{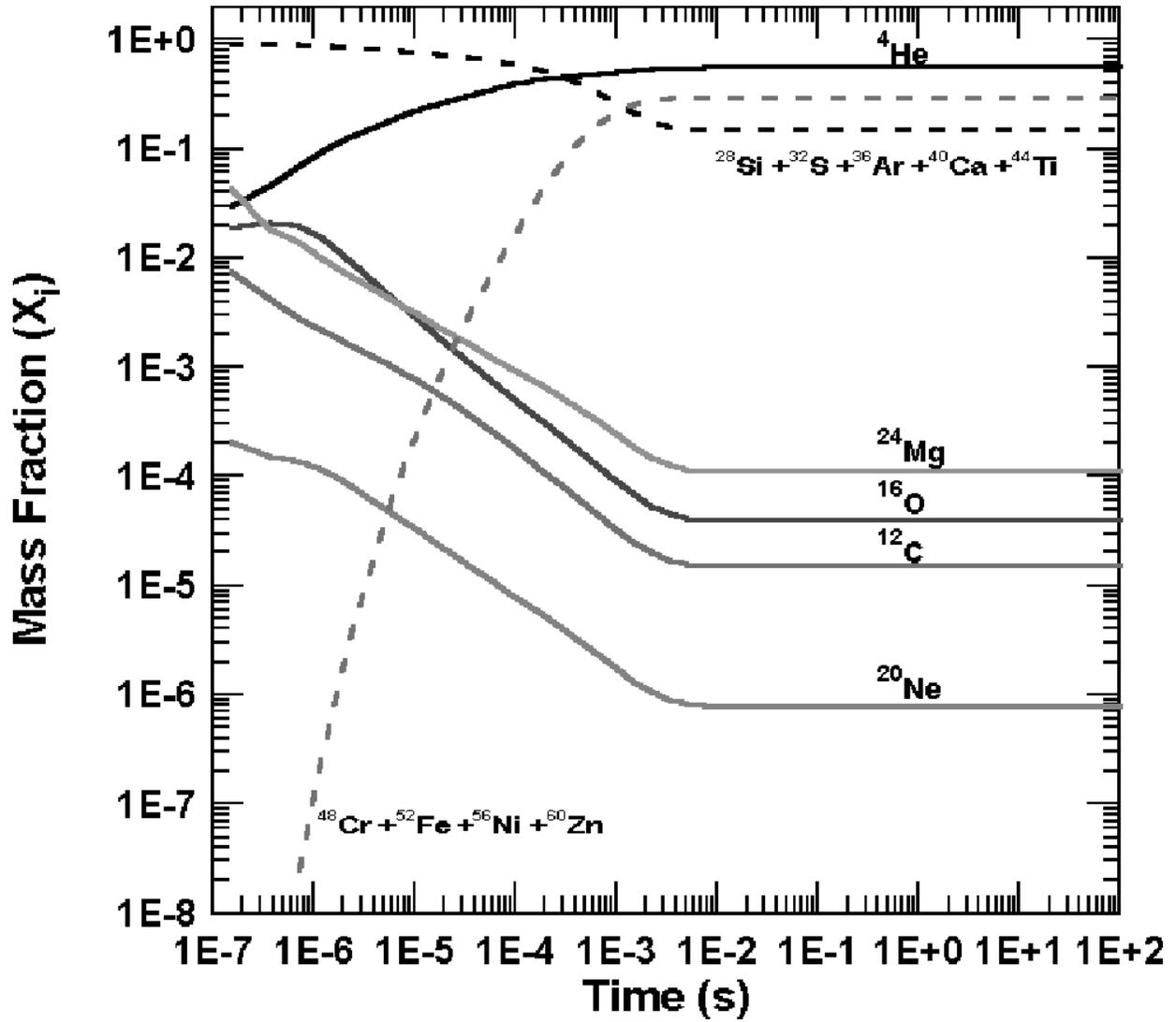}
\caption{The same as Figure 8 but for abundances in the hydrostatic calculation
(14-isotope chain).
\label{fig9}}
 \end{figure}

\clearpage

\begin{figure}
\plotone{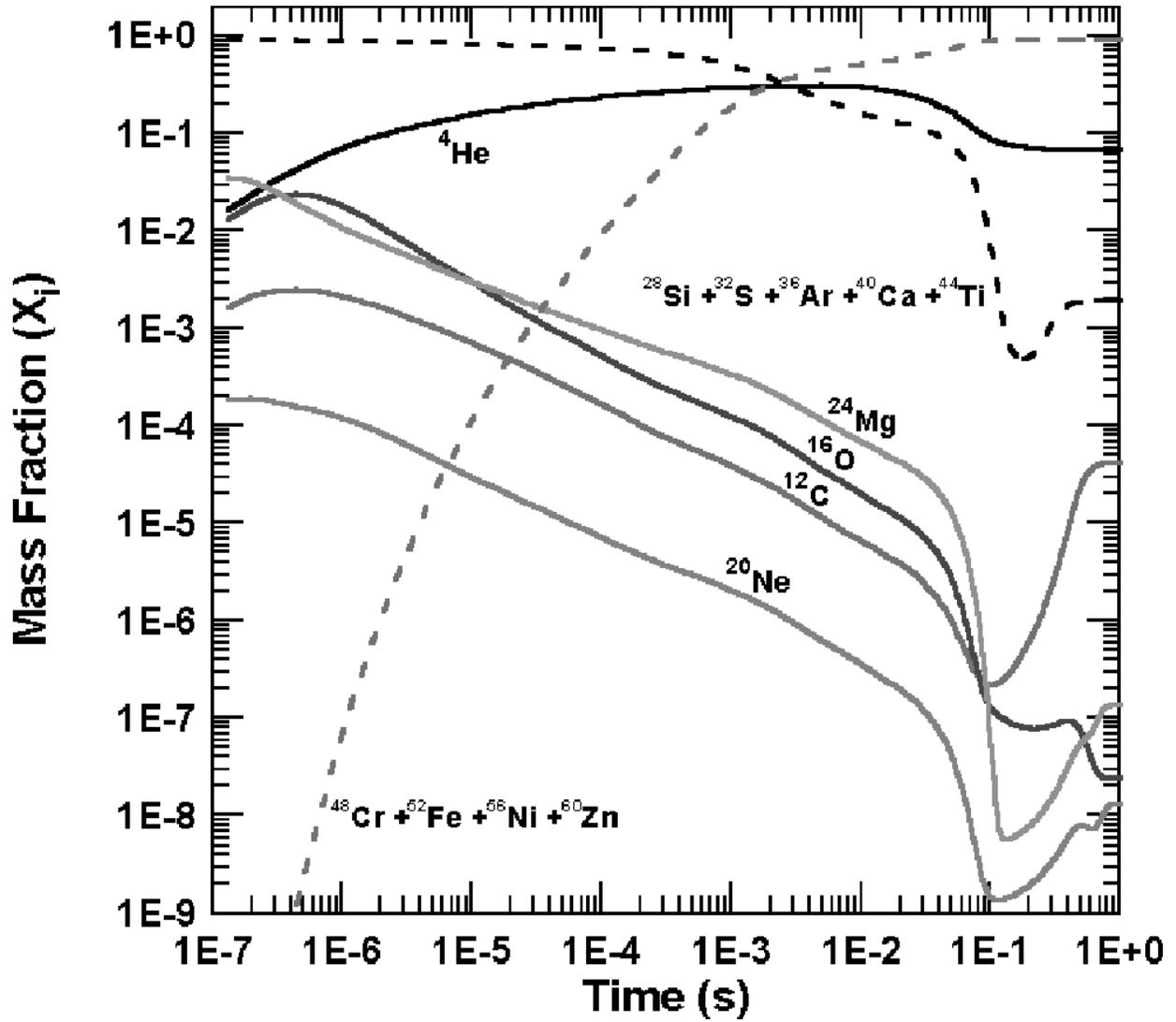}
\caption{The same as in Figure 9 but for the hydrodynamic silicon combustion.
Final products are 
typical of an $\alpha$-rich freeze out (14-isotope chain).
\label{fig10}}
 \end{figure}

\clearpage

\begin{figure}
\plotone{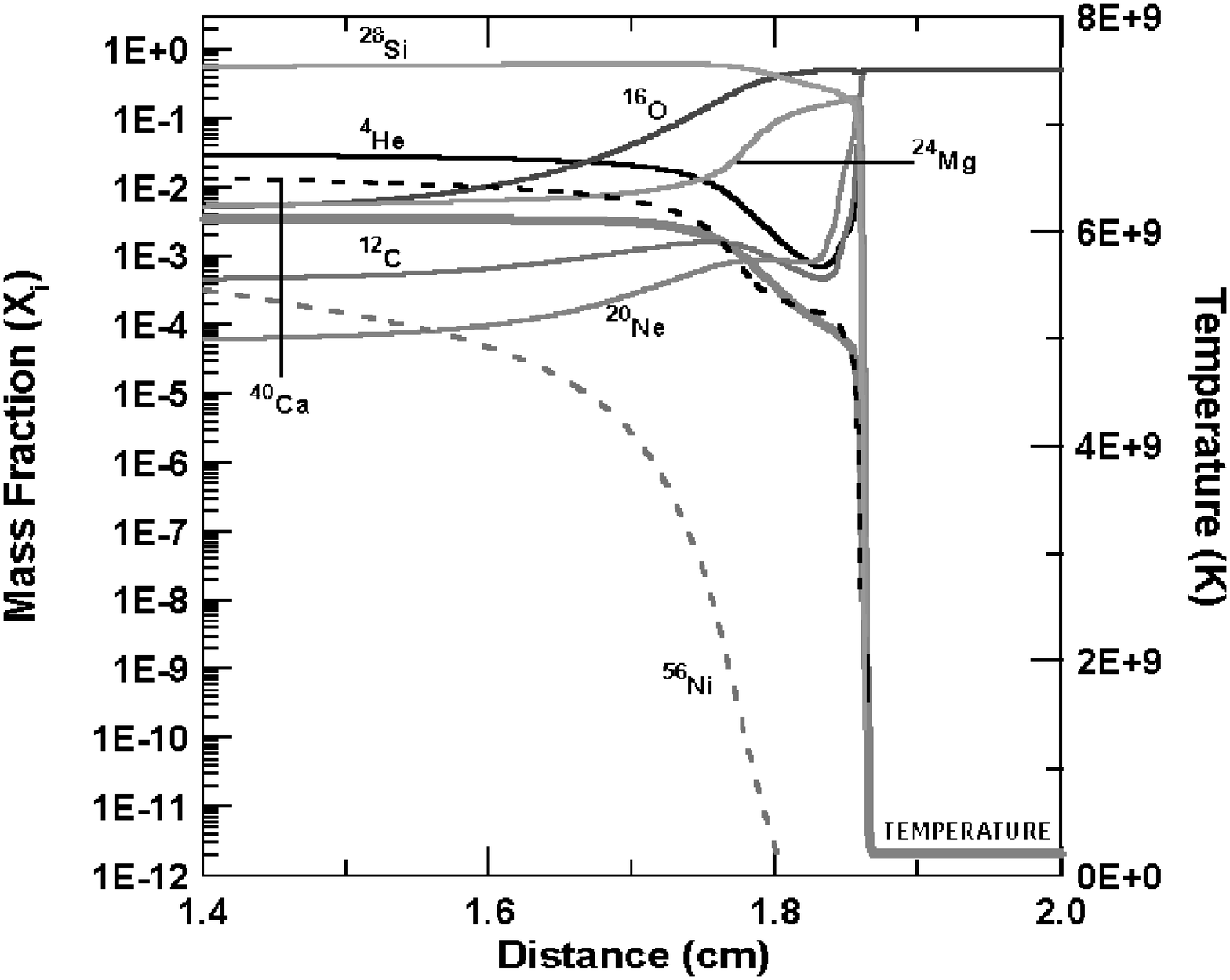}
\caption{Temperature and chemical profiles corresponding to a
nuclear flame moving to the right through a medium of density
$\rho=1.26\times 10^8$~g.cm$^{-3}$~composed of equal parts carbon and oxygen.
Again, the feedback between nuclear rates and temperature was adequately
handled during the integration, and the relaxation to  NSE was smooth
(86-isotope network).
\label{fig11}}
 \end{figure}

\clearpage

\begin{deluxetable}{crrrrr|rrrrrr}
\tabletypesize{\scriptsize}
\tablecaption{Abundances during the adiabatic expansion from NSE at
$\rho_0=10^9$~g.cm$^{-3}$~starting from the NSE temperature 
$T_0=8.03~10^9$~K. \label{tbl-1}}
\tablewidth{0pt}
\tablehead{
\colhead{Isot.} & $T_9: 8.03$& 4 &
3.0 &
2.0  & 0.5 & \colhead{Isot.} & $T_9: 8.03$&
4.0&3.0&
2.0&0.5
}
\startdata
n &5.7(-5)&$4.8(-12)$&3.0(-16)&8.6(-25)&4.0(-85)&$^{38}\mathrm{Ca}$&2.5(-7) &3.1(-14) &2.2(-12) &1.1(-8)&1.3(-8) \\

p &2.1(-2) &1.7(-3) &5.2(-4) &2.6(-4) &1.6(-6) &$^{39}\mathrm{Ca}$ &2.3(-5) &7.1(-12) &2.2(-12)  &2.6(-11) &2.6(-11)\\

$^4\mathrm{He}$ &1.95(-1) &3.4(-3) &4.5(-3) &4.4(-3) &4.4(-3) &$^{40}\mathrm{Ca}$  &5.6(-3) &6.6(-8) &1.3(-7) &4.6(-7) &4.6(-7)\\

$^{12}C$ &2.5(-5) &1(-8) &1.1(-7) &4.1(-7) &5.1(-7) &$^{41}\mathrm{Ca}$  &1.3(-3) &1.7(-11) &1.9(-12) &1.5(-18) &1.3(-21)\\

$^{16}O$ &5.3(-5) &1.7(-9) &3.2(-9) &1.1(-9)&3.7(-10)  &$^{42}\mathrm{Ca}$  &7.1(-4) &6.4(-13) &7.6(-15)  &1.2(-20) &6.8(-27)\\

$^{20}\mathrm{Ne}$ &3(-6) &3.3(-11)  &1.3(-10)  &1.6(-11)&4.7(-12) &$^{41}\mathrm{Sc}$ &1.5(-5) &4(-12) &4.8(-12) &4.2(-11) &1.8(-10)\\

$^{21}\mathrm{Ne}$ &4.3(-8) &6.2(-17) &2.3(-20)  &8.8(-30) &2(-44)&$^{42}\mathrm{Sc}$   &5.5(-5) &7.9(-13) &6.6(-13)  &3.5(-16) &3(-19)\\

$^{22}\mathrm {Ne}$ &5(-9) &1.4(-19) &2.5(-22)&7.5(-32) &1.3(-65)& $^{43}\mathrm{Sc}$  &6.2(-4) &4.6(-12) &1.2(-12) &6.9(-15) &1.9(-17)\\

$^{21}\mathrm{Na}$ &4(-8) &7.7(-14)&6.2(-13) &1.9(-12)&5.1(-12) &$^{42}\mathrm{Ti}$ &4.8(-8) &1.1(-14) &8.8(-14) &3.3(-10) &8(-10)\\

$^{22}\mathrm{Na}$ &7.2(-8) &9.4(-15)&3.1(-16) &4.6(-22) &2(-40)  &$^{43}\mathrm{Ti}$  &1.8(-6) &2.6(-13)&4.8(-12)  &8.3(-12) &8.3(-12)\\

$^{23}\mathrm{Na}$ &4.1(-7) &3.1(-14) &5.2(-14) &1.1(-16) &4.3(-25) &$^{44}\mathrm{Ti}$&1.2(-3) &1.3(-8) &6(-8) &7.9(-8) &5.5(-8)\\

$^{22}\mathrm{Mg}$ & 1.3(-9) &4.1(-14) &1.4(-11) &4.9(-8) &5.9(-8) &$^{45}\mathrm{Ti}$&1.7(-3) &2.1(-10) &1.4(-12) &5.3(-20) &3.9(-24)\\

$^{23}\mathrm{Mg}$ &1.4(-7) &7.9(-12)  &7.8(-11) &7.9(-11)&7.9(-11) &$^{46}\mathrm{Ti}$ &4.6(-3) &5.4(-10) &3.9(-13) &2.4(-19) &1.9(-23)\\

$^{24}\mathrm{Mg}$ &1.3(-4) &7.5(-11)  &4.2(-10) &1.1(-11) &9.1(-12)  &$^{45}\mathrm{Va}$  &5.8(-6) &7(-12) &3.3(-11) &3(-10) &1.5(-9)\\

$^{25}\mathrm{Mg}$ &3.9(-6) &7.3(-16) &3.7(-19) &2.6(-29) &1.3(-40) &$^{46}\mathrm{Va}$  &1.3(-4) &7.1(-11) &1(-11) &2.4(-15) &2.7(-18)\\

$^{26}\mathrm{Mg}$ &9.1(-7) &1.2(-17) &1.1(-19) &8.4(-29) &2.2(-51) &$^{47}\mathrm{Va}$  &4.8((-3) &7.8(-9) &1.7(-10) &6.5(-13) &1.1(-13)\\

$^{25}\mathrm{Al}$ &1.1(-6) &1.1(-13) &1.5(-12) &6.7(-13) &1.2(-11) &$^{46}\mathrm{Cr}$  &2.9(-8) &2(-13)&2.2(-11) &1.5(-7) &1.8(-7)\\

$^{26}\mathrm{Al}$ &4.2(-6) &4.7(-14)&1.7(-15)&2.9(-21) &1.8(-32) &$^{47}\mathrm{Cr}$  &7.9(-6) &5.7(-11) &2.5(-10) &2.6(-10) &2.6(-10)\\

$^{27}\mathrm{Al}$ &4.5(-6) &6.9(-13) &3.6(-12) &7.8(-15) &2.8(-17) &$^{48}\mathrm{Cr}$  &4(-3) &4.3(-6) &2.2(-7) &2(-7) &5.7(-8)\\

$^{26}\mathrm{Si}$ &3.1(-8) &5.3(-14) &2.3(-11) &3.4(-8) &3.5(-8) &$^{49}\mathrm{Cr}$  &1.2(-2) &6.9(-7) &2.1(-9) &2.8(-16) &2.6(-19)\\

$^{27}\mathrm{Si}$ &5.3(-6) &2(-11)&2(-10) &2(-10) &2(-10) &$^{50}\mathrm{Cr}$  &5.2(-2) &2(-6) &1.6(-12) &5.7(-20) &7.3(-24)\\

$^{28}\mathrm{Si}$ &8.9(-3) &5(-10) &3.3(-9) &3.6(-11) &1(-12) &$^{49}\mathrm{Mn}$  &2.2(-5) &3.9(-9) &3.4(-10) &5.4(-9) &3.5(-9)\\

$^{29}\mathrm{Si}$ &6(-4) &4.5(-14) &9.7(-19) &1.4(-25) &1.7(-34) &$^{50}\mathrm{Mn}$  &7.5(-4) &1.3(-7) &5.7(-9) &1.2(-12) &7.1(-15)\\

$^{30}\mathrm{Si}$ &1.7(-4) &5.3(-16) &1.1(-17) &1.3(-26) &7.1(-42) &$^{51}\mathrm{Mn}$  &3.1(-2) &2.1(-5) &6.2(-10)&1.8(-13) &1.3(-14)\\

$^{29}\mathrm{P}$ &6.5(-5) &9.8(-13) &1.8(-11) &1.3(-11) &2.5(-12) &$^{50}\mathrm{Fe}$  &1.3(-7) &2.5(-11) &2.4(-11) &3.3(-7) &4.8(-7)\\

$^{30}\mathrm{P}$ &2.2(-4) &3(-13) &2.7(-14)&6(-19) &7.1(-27) &$^{51}\mathrm{Fe}$  &3.6(-5) &7.3(-8) &8.6(-8) &9.4(-8) &9.4(-8)\\

$^{31}\mathrm{P}$ &9.2(-4) &6.2(-13) &2.8(-12) &1.4(-15) &3.1(-19) &$^{52}\mathrm{Fe}$  &1.3(-2) &2.1(-3) &8.9(-6) &7.6(-6) &7.5(-6)\\

$^{30}\mathrm{S}$ &9.2(-7) &5.6(-14)&1.5(-11) &1.6(-8) &1.7(-8) &$^{53}\mathrm{Fe}$  &5.1(-2) &4.1(-4) &1.6(-5) &7.7(-12) &6.8(-14)\\

$^{31}\mathrm{S}$ &4.6(-5) &3.3(-12) &2.4(-11) &2.4(-11) &2.4(-11) &$^{54}\mathrm{Fe}$  &2.7(-1) &3(-3) &5.5(-7) &3.6(-16) &8.4(-20)\\

$^{32}\mathrm{S}$ &8.8(-3) &1.2(-9) &1.1(-8) &2.7(-10) &2.6(-12) &$^{53}\mathrm{Co}$  &2.5(-5) &4.1(-7) &2(-9) &1.3(-8) &1.3(-8)\\

$^{33}\mathrm{S}$ &1.5(-3) &3.6(-13) &1.8(-15) &1(-23) &1.4(-30) &$^{54}\mathrm{Co}$  &2.1(-3) &4.8(-5) &2.3(-5) &1(-8) &3.4(-10)\\

$^{34}\mathrm{S}$ &7.6(-4) &2.1(-14) &1.5(-15) &2.5(-23) &1(-34) &$^{55}\mathrm{Co}$  &9.6(-2) &1.9(-2) &9.9(-5) &3.7(-10) &8.4(-12)\\

$^{33}\mathrm{Cl}$ &6.5(-5) &1.2(-12) &2(-11) &1.3(-11) &1.8(-12) &$^{54}\mathrm{Ni}$  &7.8(-8) &1.3(-9) &5.2(-11) &1.2(-7) &2.3(-7)\\

$^{34}\mathrm{Cl}$ &3.2(-4) &6.1(-13) &7.7(-14) &2.1(-18) &1.1(-25) &$^{55}\mathrm{Ni}$  &5.3(-5) &9.1(-6) &8.9(-5) &1.3(-4) &1.3(-4)\\

$^{35}\mathrm{Cl}$ &1.9(-3) &3.1(-14) &2.1(-11) &2.5(-14) &2.8(-17) &$^{56}\mathrm{Ni}$  &2.3(-2) &9.18(-1) &9.65(-1) &9.63(-1) &9.63(-1)\\

$^{34}\mathrm{Ar}$ &8(-7) &7.4(-14) &2.3(-11) &3.6(-8) &3.6(-8) &$^{57}\mathrm{Ni}$  &5.5(-2) &3.4(-2) &2.7(-2) &1.8(-2) &1(-2)\\

$^{35}\mathrm{Ar}$ &4.5(-5) &3.4(-12) &3(-11) &3.1(-11) &3.1(-11) &$^{58}\mathrm{Ni}$  &1.14(-1) &1.8(-2) &1.4(-3)&1.6(-8) &7.2(-10)\\

$^{36}\mathrm{Ar}$ &6.3(-3) &2.2(-9) &8.2(-9) &2.94(-9) &7.9(-10) &$^{57}\mathrm{Cu}$  &2.2(-5) &9.5(-6) &3.9(-6) &4.7(-6) &5.3(-6)\\

$^{37}\mathrm{Ar}$ &1.3(-3) &1(-12) &2.6(-15) &1.9(-23) &2.74(-29) &$^{58}\mathrm{Cu}$  &5(-4) &6.2(-5) &1.4(-4) &6.2(-3) &6.8(-3)\\

$^{38}\mathrm{Ar}$ &1.1(-3) &1.9(-13) &5.4(-15) &8(-22) &6.5(-30) &$^{59}\mathrm{Cu}$  &5.4(-3) &6(-4) &2.4(-4) &5.8(-7) &1.7(-7)\\

$^{37}\mathrm{K}$ &2.8(-5) &6.8(-13) &3(-12) &1.2(-11) &7.1(-11) &$^{58}\mathrm{Zn}$  &1.2(-8) &6(-10) &4.5(-10) &1.4(-8) &2.7(-9)\\

$^{38}\mathrm{K}$ &2.3(-4) &2.3(-12) &1.6(-13) &7(-18) &1.1(-22) &$^{59}\mathrm{Zn}$  &1.8(-6) &8.7(-8) &6.7(-7) &2.8(-3) &1.02(-2)\\

$^{39}\mathrm{K}$ &2.3(-4) &3.1(-11) &7.9(-11) &8.5(-13) &1.1(-13) &$^{60}\mathrm{Zn}$  &1.3(-4) &1.5(-4) &1.7(-3) &4.9(-3) &4.9(-3)\\

\enddata
\end{deluxetable}

\clearpage

\begin{deluxetable}{crrrrrr|rrrrc}
\tabletypesize{\scriptsize}
\tablecaption{Five more abundant nuclei at NSE and at the freeze-out temperature
for the models considered in Sec. 3.1.  
The NSE temperature and mass fractions of these nuclei are given in the left 
side of the Table,  
as a function of the initial composition of the fuel. The approximate freeze out temperature and the corresponding abundances are also given in the right part 
of the Table.
A specie is considered burned once its  
 mass fraction differs from its final stable value in less than 2\%. 
Temperatures are in units of $10^9$~K. 
\label{tbl-1}}
\tablewidth{0pt}
\tablehead{
\colhead{Fuel} &\colhead{} &\colhead{} &\colhead{NSE}&\colhead{}&\colhead{}&
\colhead{}&\colhead{}&
\colhead{FREEZE}&\colhead{OUT}&\colhead{}&\colhead{}}
 \startdata
 &Ash  &$^{54}$Fe&$^{4}$He&$^{58}$Ni&$^{55}$Co&$^{57}$Ni&
$^{56}$Ni&$^{59}$Zn&$^{57}$Ni&$^{58}$Cu&$^{60}$Zn\\
CO&$X_k$&$2.7(-1)$&$1.95(-1)$&$1.14(-1)$&
$9.55(-2)$&$5.46(-2)$&
$9.63(-1)$&$1.02(-2)$&$1.01(-2)$&
$6.82(-3)$&$4.93(-3)$\\
 &T$_9$&8.03&8.03&8.03&8.03&8.03&3.20&1.31&1.37&1.35&2.10\\
\hline
&Ash &$^{54}$Fe&$^{4}$He&$^{50}$Cr&$^{58}$Ni&$^{34}$S&
  $^{54}$Fe&$^{58}$Ni&$^{50}$Cr&$^{4}$He&$^{46}$Ti\\
CONe&$X_k$&$4.1(-1)$&$1.99(-1)$&$1.59(-1)$&
$1.04(-1)$&$3.08(-2)$&
$7.25(-1)$&$2.12(-1)$&$5.58(-2)$&
$9.49(-3)$&$3.11(-3)$\\
 &T$_9$&8.51&8.51&8.51&8.51&8.51&5.52&4.77&4.75&4.45&4.63\\
\hline
&Ash&$^{56}$Ni&$^{55}$Co&$^{54}$Fe&$^{57}$Ni&$^{58}$Ni&
  $^{56}$Ni&$^{57}$Ni&$^{59}$Zn&$^{58}$Cu&$^{4}$He\\
He&$X_k$&$6.2(-1)$&$1.1(-1)$&$9.3(-2)$&
$5.0(-2)$&$3.8(-2)$&
$9.7(-1)$&$8.2(-3)$&$6.5(-3)$&
$4.3(-3)$&$1.7(-3)$\\
 &T$_9$&4.30&4.32&4.32&4.32&4.32&2.95&1.14&1.11&1.23&2.43\\
\enddata
\end{deluxetable}


\clearpage
\begin{thebibliography}{}

\bibitem[1999]{a99} Angulo C., et al., 1999, Nuc. Phys., A656,3

\bibitem[1996]{a96} Arnett,D., 1996, Supernovae and Nucleosynthesis,
  (Princenton University Press)

\bibitem[1969]{at69} Arnett, D.W., Truran, J.W., 1969, ApJ, 157, 339.

\bibitem[1988]{cf88} Caughlan, G.R., Fowler, W.A., 1988, Atomic Data and
Nuclear Data Tables 40, 283.

\bibitem[1968]{c68} Clayton, D.D., 1968, Principles of Stellar Evolution
and Nucleosynthesis. (Chicago: Univ. Chicago Press).

\bibitem[1998]{bgs98} Bravo, E., Garc\'\i a-Senz, 1999, MNRAS, 307, 984

\bibitem[2003]{gsc03} Garc\'\i a-Senz, D., Cabez\'on, R.M., 2003, Nuc. Phys.A,
718, 566c.

\bibitem[1996]{ht96} Hix, W.R., Thielemann, F.-K., 1996, ApJ, 406, 869.

\bibitem[1997]{kow97} Khokhlov, A.M., Oran, E.S., Wheeler, J.C., 1997, ApJ, 478, 678

\bibitem[1998]{mkc98} Meyer, B.S., Krishnan, T.D., Clayton, D.D., 1998, ApJ,
498, 808.

\bibitem[1986]{mn86} Mochkovitch, R., Nomoto, K., 1986, A\&A, 154, 115

\bibitem[1986]{m86} M\"uller, E., 1986, A\&A, 162, 103.

\bibitem[1987]{oi87} Ogata, S., Ichimaru, S., 1987, Phys. Rev. A, 36, 5451

\bibitem[1987]{paa87} Prantzos, N, Arnould, M., Arcoragi, J.P., 1987, ApJ, 315,
209.

\bibitem[2000]{rt00} Rauscher T., Thielemann, F.-K., 2000, Atomic Data and
Nuclear Data Tables 75, 1.

\bibitem[2001]{rhhw01} Rauscher, T., Heger, A., Hoffman, R.D., Woosley, S.E.,
2001, Nuc. Phys.A, 688, 193c-196c.

\bibitem[1999]{t92} Timmes, F.X., 1999, ApJS, 124, 241.

\bibitem[1992]{tw92} Timmes, F.X., Woosley, S.E., 1992, ApJ, 396, 649.

\bibitem[2000]{thw00} Timmes, F.X., Hoffman, R.D., Woosley, S.E., 2000,
ApJS, 129, 377


\bibitem[1973]{wac73} Woosley, S.E., Arnett, W.D., Clayton, D.D., 1973, ApJS,
26, 231.

\bibitem[1989]{ys89} Yakovlev, D.G., Shalybkov, D.A., 1989, Soviet Sci.Rev.E, 7,311

\end{thebibliography}
\end{document}